# A regional model for estimating the aboveground carbon density of Borneo's tropical forests from airborne laser scanning

Tommaso Jucker[1,2*], Gregory P. Asner[3], Michele Dalponte[4], Philip Brodrick[3], Christopher D. Philipson[5,6], Nick Vaughn[3], Craig Brelsford[7], David F.R.P. Burslem[8], Nicholas J. Deere[9], Robert M. Ewers[10], Jakub Kvasnica[11], Simon L. Lewis[12,13], Yadvinder Malhi[14], Sol Milne[7], Reuben Nilus[15], Marion Pfeifer[16], Oliver Phillips[12], Lan Qie[10,12], Nathan Renneboog[17], Glen Reynolds[18], Terhi Riutta[14], Matthew J. Struebig[9], Martin Svátek[11], Yit Arn Teh[7], Edgar C. Turner[19] and David A. Coomes[1*]

[*]Authors contributed equally to the writing of this paper

[1]Forest Ecology and Conservation group, Department of Plant Sciences, University of Cambridge, UK (dac18@cam.ac.uk)

[2]CSIRO Land and Water Flagship, Private Bag 5, P.O. Wembley, Western Australia 6913, Australia (tommaso.jucker@csiro.au)

[3]Department of Global Ecology, Carnegie Institution for Science, 260 Panama Street, Stanford, CA 94305 USA (gpa@carnegiescience.edu; pbrodrick@carnegiescience.edu; nvaughn@carnegiescience.edu)

[4]Department of Sustainable Agro-ecosystems and Bioresources, Research and Innovation Centre, Fondazione E. Mach, Via E. Mach 1, 38010 San Michele all'Adige, Italy (michele.dalponte@fmach.it)

[5]Department of Environmental Systems Science, ETH Zürich, Universitätstrasse 16, 8092 Zürich, Switzerland (christopher.philipson@usys.ethz.ch)

[6]Centre for Environmental Change and Human Resilience, University of Dundee, Dundee, UK

[7]Department of Biosciences, Viikki Plant Science Center (ViPS), University of Helsinki, 00014, Helsinki, Finland (craig.brelsford@helsinki.fi)

[8]School of Biological Sciences, University of Aberdeen, Cruickshank Building, St Machar Drive, Aberdeen, UK (yateh@abdn.ac.uk, d.burslem@abdn.ac.uk, s.milne@abdn.ac.uk)

[9]Durrell Institute of Conservation and Ecology (DICE), School of Anthropology and Conservation, University of Kent, Canterbury, UK (njd21@kent.ac.uk; m.j.struebig@kent.ac.uk)

[10]Imperial College London, Silwood Park Campus, Buckhusrt Road, Ascot SL5 7PY, UK (r.ewers@imperial.ac.uk)

[11]Faculty of Forestry and Wood Technology, Department of Forest Botany, Dendrology and Geobiocoenology, Mendel University, Brno, Czech Republic (msvatek@centrum.cz; jakub.kvasnica@mendelu.cz)

[12]School of Geography, University of Leeds, Leeds LS2 9JT, UK (s.l.lewis@leeds.ac.uk; o.phillips@leeds.ac.uk; l.qie@leeds.ac.uk)

[13]Department of Geography, University College London, London, UK

[14]Environmental Change Institute, School of Geography and the Environment, University of Oxford, Oxford, UK (yadvinder.malhi@ouce.ox.ac.uk; terhi.riutta@ouce.ox.ac.uk)

[15]Forest Research Centre, Sabah Forestry Department, P.O. Box 1407, 90715 Sandakan, Sabah, Malaysia (reuben.nilus@sabah.gov.my)

[16]School of Biology, Newcastle University, Newcastle NE17RU, UK (marion.pfeifer@newcastle.ac.uk)

[17]Permian Global, Savoy Hill House, 7-10 Savoy Hill, London WC2R 0BU, UK (nathan.renneboog@permianglobal.com)

[18]South East Asia Rainforest Research Partnership (SEARRP), Danum Valley Field Centre, PO Box 60282, 91112 Lahad Datu, Sabah, Malaysia (glen.searrp@icloud.com)

[19]Department of Zoology, University of Cambridge, Downing Street, Cambridge, CB2 3EJ, UK (ect23@cam.ac.uk)

**Correspondence**: David A. Coomes; Forest Ecology and Conservation group, Department of Plant Sciences, University of Cambridge, Downing Street, Cambridge CB2 3EA, UK; Email: dac18@cam.ac.uk); Phone: +44 (0)1223 333911; Fax: +44 (0)1223 333953.

**Author contribution**: D.A.C. and Y.A.T. coordinated the NERC airborne campaign, while G.P.A. led the CAO airborne surveys of Sabah. T.J. and D.A.C. designed the study, with input from G.P.A. and P.B.; M.D., T.J., P.B. and N.V. processed the airborne imagery, while other authors contributed field data; T.J. and D.A.C. analysed the data, with input from G.P.A., P.B. and C.D.P.; T.J. and D.A.C. wrote the first draft of the manuscript, with all other authors contributing to revisions.

**Running head**: Remote sensing of tropical forest carbon

**Abstract**

Borneo contains some of the world's most biodiverse and carbon dense tropical forest, but this 750,000-km$^2$ island has lost 62% of its old-growth forests within the last 40 years. Efforts to protect and restore the remaining forests of Borneo hinge on recognising the ecosystem services they provide, including their ability to store and sequester carbon. Airborne Laser Scanning (ALS) is a remote sensing technology that allows forest structural properties to be captured in great detail across vast geographic areas. In recent years ALS has been integrated into state-wide assessment of forest carbon in Neotropical and African regions, but not yet in Asia. For this to happen new regional models need to be developed for estimating carbon stocks from ALS in tropical Asia, as the forests of this region are structurally and compositionally distinct from those found elsewhere in the tropics. By combining ALS imagery with data from 173 permanent forest plots spanning the lowland rain forests of Sabah, on the island of Borneo, we develop a simple-yet-general model for estimating forest carbon stocks using ALS-derived canopy height and canopy cover as input metrics. An advanced feature of this new model is the propagation of uncertainty in both ALS- and ground-based data, allowing uncertainty in hectare-scale estimates of carbon stocks to be quantified robustly. We show that the model effectively captures variation in aboveground carbons stocks across extreme disturbance gradients spanning tall dipterocarp forests and heavily logged regions, and clearly outperforms existing ALS-based models calibrated for the tropics, as well as currently available satellite-derived products. Our model provides a simple, generalised and effective approach for mapping forest carbon stocks in Borneo, providing a key tool to support the protection and restoration of its tropical forests.

# 1    Introduction

Forests are an important part of the global carbon cycle (Pan et al., 2011), storing and sequestering more carbon than any other ecosystem (Gibbs et al., 2007). Estimates of tropical deforestation rates vary, but roughly 230 million hectares of forest were lost per year between 2000 and 2012, and an additional 30% were degraded by logging or fire (Asner et al., 2009; Hansen et al., 2013). Forest degradation and deforestation causes substantial releases of greenhouse gases to the atmosphere – about 1-2 billion tonnes of carbon per year – which equates to about 10% of global emissions (Baccini et al., 2012). Even if nations de-carbonise their energy supply chains within agreed schedules, a rise of 2°C in mean annual temperature is unavoidable unless 300 million hectares of degraded tropical forests are protected, and land unsuitable for agriculture is reforested (Houghton et al., 2015). Signatories to the Paris agreement, brokered at COP21 in 2015, are now committed to reducing emissions from tropical deforestation and forest degradation (i.e., REDD+; Agrawal et al., 2011), whilst recognising that these forests also harbour rich biodiversity and support livelihoods for around a billion people (Vira et al., 2015).

Accurate monitoring of forest carbon stocks underpins these initiatives to generate carbon credits through REDD+ and similar forest conservation and climate change mitigation programs (Agrawal et al., 2011). Airborne Laser Scanning (ALS) has shown particular promise in this regard, because it generates high resolution maps of forest structure from which aboveground carbon density (*ACD*) can be estimated (e.g., Asner et al., 2010; Lefsky et al., 1999; Nelson et al., 1988; Popescu et al., 2011; Wulder et al., 2012). The principle of ALS is that laser pulses are emitted downwards from an aircraft, and a sensor records the time it takes for individual beams to strike a surface (e.g., leaves, branches or the ground) and bounce back to the emitting source, thereby precisely measuring the distance between the object and the airborne platform. Divergence of the beam means it is wider than leaves and allows penetration into the canopy, resulting is a 3D point cloud that captures the vertical structure of the forest. By far the most common approach to using ALS data for estimating forest carbon stocks involves developing statistical models relating *ACD* estimates obtained from permanent field plots to summary statistics derived from the ALS point cloud, such as the mean height of returns or their skew (Zolkos et al., 2013). These "area-based" approaches were first used for mapping structural attributes of complex multi-layered forests in the early 2000s (Drake et al., 2002; Lefsky et al., 2002), and have since been applied to carbon mapping in several tropical regions (Asner et al., 2014, 2010; Baraloto et al., 2012; Jubanski et al., 2013; Longo et al., 2016; Réjou-Méchain et al., 2015; Vaglio Laurin et al., 2014).

This paper develops a statistical model for mapping forest carbon, and its uncertainty, in Southeast Asian forests. We work with ALS and plot data collected in the Malaysian state of Sabah, on the north-eastern end of the island of Borneo (Fig. 1), which is an important testbed for international efforts to protect and restore tropical forests. Borneo lost more than 62% of its old-growth forest in just 40 years as a result of heavy logging, and subsequent establishment of oil palm and forestry plantations (Gaveau et al., 2016, 2014). Sabah lost its forests at an even faster rate in this period (Osman et al., 2012), and because these forests are amongst the most carbon dense in the tropics, carbon loss has been considerable (Carlson et al., 2012a, 2012b; Slik et al., 2010). In response to past and ongoing forest losses, new initiatives have been undertaken to incentivize forest protection, including REDD+. The Sabah state government has also recently made efforts to become a regional leader in forest conservation and sustainable management, via initiatives such as the Heart of Borneo Project (http://www.heartofborneo.org/). These efforts require mapping and monitoring of forest attributes, such as carbon stocks, to target high-value conservation forests and/or to develop plans for forest restoration (Ioki et al., 2014).

[5]

The approach we develop builds upon the work of Asner & Mascaro (2014). They proposed a generalised approach for estimating *ACD* using a single ALS metric – the mean top-of-canopy height (*TCH*, in m) – and minimal field data inputs. The method relates *ACD* to *TCH*, stand basal area (*BA*; in $m^2$ $ha^{-1}$) and the community-weighted mean wood density (*WD*; in g $cm^{-3}$) over a prescribed area of forest such as one hectare, as follows:

$$ACD_{General} = 3.836 \times TCH^{0.281} \times BA^{0.972} \times WD^{1.376} \tag{1}$$

Asner & Mascaro (2014) demonstrated that tropical forests from 14 regions differ greatly in structure. Remarkably, they found that a generalised power-law relationship could be fitted that transcended these contrasting forests types, once regional differences in structure were incorporated as sub-models relating *BA* and *WD* to *TCH*. However, this general model may generate systematic errors in *ACD* estimates if applied to regions outside the calibration range, and Asner & Mascaro (2014) make clear that regional models should be obtained where possible. Since South East Asian rainforests were not among the 14 regions used to calibrate the general model, and are distinct from Neotropical and Afrotropical forests in structure and taxonomy (Banin et al., 2012), new regional models are needed before Borneo's forest carbon stocks can be surveyed using ALS. Central to the robust estimation of *ACD* using ALS data is identifying a metric which captures variation in basal area among stands. Asner & Mascaro's (2014) power-law model rests on an assumption that basal area is closely related to top-of-canopy height, an assumption supported in some studies, but not in others (Coomes et al., 2017; Duncanson et al., 2015; Spriggs, 2015). The dominance of Asian lowland rainforests by dipterocarp species make them structurally unique (Banin et al., 2012; Feldpausch et al., 2011;

Ghazoul, 2016) and gives rise to greater aboveground carbon densities than found in any other tropical region (Avitabile et al., 2016; Sullivan et al., 2017), highlighting the need for new ALS-based carbon estimation models for this region.

Here, we develop and calibrate a regional ALS-based model of aboveground carbon density (*ACD*, in Mg C ha$^{-1}$) to support forest carbon mapping of Borneo's forests. We bring together ALS data with estimates of *ACD* from a total of 173 permanent forests plots spanning the major lowland dipterocarp forest types and disturbance gradients found in Borneo. Using a regression framework we combined field estimates of *ACD* with simple ALS metrics to generate models for predicting carbon stocks and their uncertainty for Borneo's forests at hectare resolution. We then compared the accuracy of these models against that of existing ALS-derived equations of *ACD* developed for the tropics (Asner and Mascaro, 2014), as well as satellite-based carbon maps of the region (Avitabile et al., 2016; Pfeifer et al., 2016).

## 2       Methods

### 2.1      Study region

The study was conducted in Sabah, a Malaysian state in northern Borneo (Fig. 1a). Mean daily temperature is 26.7 °C (Walsh and Newbery, 1999) and annual rainfall is 2600-3000 mm (Kumagai and Porporato, 2012). Severe droughts linked to El Niño events occur about once every ten years (Malhi & Wright, 2004). Sabah supports a wide range of forests types, including dipterocarp forests in the lowlands that are among the tallest in the tropics (Fig. 1b; Banin et al., 2012).

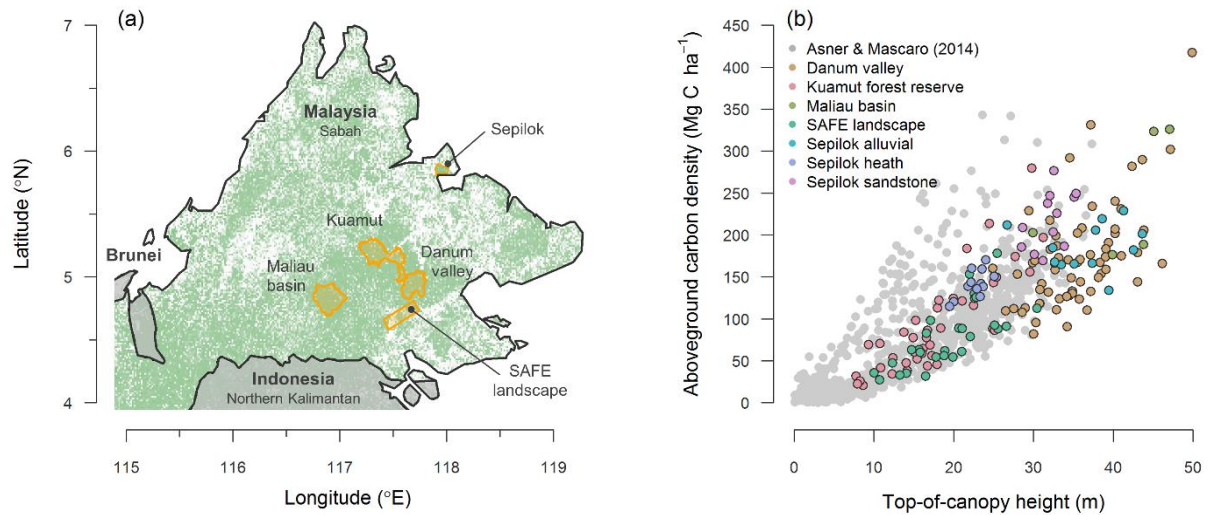

**Fig. 1**: Panel (**a**) shows the location of the Sepilok and Kuamut Forest Reserves, the Danum Valley and Maliau Basin Conservation Areas, and the SAFE landscape within Sabah (Malaysia). Green shading in the background represents forest cover at 30-m resolution in the year 2000 (Hansen et al., 2013). In panel (**b**), the relationship between field-measured aboveground carbon density and ALS-derived top-of-canopy height found across the study sites (coloured circles, $n = 173$) is compared to measurements taken mostly in the Neotropics (Asner & Mascaro 2014; grey circles, $n = 754$).

## 2.2     Permanent forest plot data

We compiled permanent forest plot data from five forested landscapes across Sabah (Fig. 1a): Sepilok Forest Reserve, Kuamut Forest Reserve, Danum Valley Conservation Area, the Stability of Altered Forest Ecosystems (SAFE) experimental forest fragmentation landscape (Ewers et al., 2011), and Maliau Basin Conservation Area. Here we provide a brief description of the permanent plot data collected at each site, which are summarized in Table 1. Additional details are provided in Appendix S1 of Supporting Information.

### 2.2.1    Sepilok Forest Reserve

The reserve is a protected area encompassing a remnant of coastal lowland old-growth tropical rainforest (Fox, 1973) and is characterized by three strongly contrasting soil types that give rise to forests that are structurally and functionally very different (Dent et al., 2006; DeWalt et al., 2006; Nilus et al., 2011): alluvial dipterocarp forest in the valleys (hereafter alluvial forests), sandstone hill dipterocarp forest on dissected hillsides and crests (hereafter sandstone forests), and heath forest on podzols associated with the dip slopes of cuestas (hereafter heath forests). There are nine permanent 4 ha forest plots within the reserve, three in each forest type. These were first established in 2000–01 and were most recently re-censused in 2013–15. All stems with a diameter ($D$, in cm) $\geq$ 10 cm were recorded and identified to species (or closest taxonomic unit). Tree height ($H$, in m) was measured for a subset of trees ($n = 718$) using a laser range finder. For the purposes of this analysis, each 4 ha plot was subdivided into 1 ha subplots, giving a total 36 plots of 1 ha in size. The corners of the plots were geolocated using a Geneq SXBlue II GPS unit, which uses satellite-based augmentation to perform differential correction and is capable of a positional accuracy of less than 2 m.

[9]

**Table 1**: Summary of permanent forest plot data collected at each study site and description of which ALS sensor was used to at each location.

| Study site | Census year | № plots | Plot size (ha) | № trees | $D_{min}$ (cm) | Height | Species ID | ALS sensor |
|---|---|---|---|---|---|---|---|---|
| Sepilok Forest Reserve | 2013–15 | 36 | 1 | 22430 | 10 | ✓ | ✓ | NERC ARF |
| Kuamut Forest Reserve | 2015–16 | 39 | 0.265[*] | 5588 | 10 | ✓ | ✓ | CAO–3 |
| Danum Valley Conservation Area | | | | | | | | |
| *CTFS plot* | 2010–16 | 45 | 1 | 215016 | 1 | ✓ | ✓ | NERC ARF |
| *CAO plots* | 2017 | 20 | 0.271[*] | 2771 | 10 | ✓ | ✓ | CAO–3 |
| SAFE landscape | | | | | | | | |
| *SAFE experiment* | 2014 | 38[†] | 0.0625 | 8444 | 1 | ✓ | | NERC ARF |
| *SAFE experiment* | 2010 | 101[†] | 0.0625 | 2517 | 10 | | | NERC ARF |
| *Riparian buffers* | 2014 | 48[†] | 0.0625 | 1472 | 10 | ✓ | | NERC ARF |
| *GEM plots* | 2014 | 4 | 1 | 1900 | 10 | ✓ | ✓ | NERC ARF |
| Maliau Basin Conservation Area | | | | | | | | |
| *SAFE experiment* | 2010 | 27[†] | 0.0625 | 894 | 10 | | | NERC ARF |
| *GEM plots* | 2014 | 2 | 1 | 905 | 10 | ✓ | ✓ | NERC ARF |

[*]Mean plot size after applying slope correction (see Sect. 2.2.2 for further details)

[†]Plots established as part of the SAFE experiment and those located along riparian buffer zones in the SAFE landscape were aggregated into spatial blocks prior to statistical analyses ($n = 27$ with a mean plot size of 0.5 ha; see Sect. 2.2.4 for further details).

## 2.2.2   Kuamut Forest Reserve

The reserve is a former logging area that is now being developed as a restoration project. Selective logging during the past 30 years has left large tracts of forest in a generally degraded condition, although the extent of this disturbance varies across the landscape. Floristically and topographically the Kuamut reserve is broadly similar to Danum Valley – with which it shares a western border – and predominantly consists of lowland dipterocarp forests. Within the forest reserve, 39 circular plots with a radius of 30 m were established in 2015–16. Coordinates for the plot centres were taken using a Garmin GPSMAP 64S device with an accuracy of $\pm10$ m. Within each plot, all stems with $D \geq 10$ cm were recorded and identified to species (or closest taxonomic unit), and $H$ was measured using a laser range finder. Because the radius of the plots was measured along the slope of the terrain (as opposed to a horizontally projected distance), we slope-corrected the area of each plot by multiplying by $\cos(\theta)$, where $\theta$ is the average slope of the plot in degrees as calculated from the digital elevation model obtained from the ALS data. The average plot size after applying this correction factor was 0.265 ha (6% less than if no slope correction had been applied).

## 2.2.3   Danum Valley Conservation Area

The site encompasses the largest remaining tract of primary lowland dipterocarp forest in Sabah. Within the protected area, we obtained data from a 50 ha permanent forest plot which was established in 2010 as part of the Centre for Tropical Forest Science (CTFS) ForestGEO network (Anderson-Teixeira et al., 2015). Here we focus on 45 ha of this plot for which all stems with $D \geq 1$ cm have been mapped and taxonomically identified (mapping of the remaining 5 ha of forest was ongoing as of January 2017). For the purposes of this study, we subdivided the mapped area into 45 1 ha plots, the coordinates of which were recorded using

the Geneq SXBlue II GPS. In addition to the 50 ha CTSF plot, we also secured data from 20

circular plots with a 30 m radius that were established across the protected area by the Carnegie

Airborne Observatory (CAO) in 2017. These plots were surveyed following the same protocols as those described previously for the plots at Kuamut in Sect. 2.2.2.

### 2.2.4   SAFE landscape and Maliau Basin Conservation Area

Plot data from three sources were acquired from the SAFE landscape and the Maliau Basin

Conservation Area: research plots established through the SAFE project, plots used to monitor

riparian buffer zones, and plots from the Global Ecosystem Monitoring (GEM) network

(http://gem.tropicalforests.ox.ac.uk). As part of the SAFE project, 166 plots of 25 × 25 m in

size were established in forested areas (Ewers et al., 2011; Pfeifer et al., 2016). Plots are organized in blocks which span a land-use intensity gradient, ranging from twice-logged forests that are currently in the early stages of secondary succession within the SAFE landscape, to relatively undisturbed old-growth forests at Maliau Basin (Ewers et al., 2011; Struebig et al.,

2013). Plots were surveyed in 2010, at which time all stems with $D \geq 10$ cm were recorded and plot coordinates were taken using a Garmin GPSMap60 device (accurate to within ±10 m). Of these 166 plots, 38 were re-surveyed in 2014, at which time all stems with $D \geq 1$ cm were recorded and tree heights were measured using a laser range finder. Using these same protocols, a further 48 plots were established in 2014 along riparian buffer zones in the SAFE landscape.

As with the SAFE project plots, riparian plots are also spatially clustered into blocks. The small size of the SAFE and riparian plots (0.0625 ha) makes them prone to high uncertainty when modelling carbon stocks from ALS (Réjou-Méchain et al., 2014), especially given the relatively low positional accuracy of the GPS coordinates. To minimize this source of error,
we chose to aggregate individual plots into blocks for all subsequent analyses ($n = 27$, with a
mean size of 0.5 ha). Lastly, we obtained data from six GEM plots – four within the SAFE
landscape and two at Maliau Basin. The GEM plots are 1 ha in size and were established in
2014. All stems with $D \geq 10$ cm were mapped, measured for height using a laser range finder,
and taxonomically identified. The corners of the plots were georeferenced using the Geneq
SXBlue II GPS.

## 2.3    Estimating aboveground carbon density and its uncertainty

Across the five study sites we compiled a total of 173 plots that together cover a cumulative
area of 116.1 ha of forest. For each of these plots we calculated aboveground carbon density
(*ACD*, in Mg C ha$^{-1}$) following the approach outlined in the *BIOMASS* package in R (R Core
Development Team, 2016; Réjou-Méchain et al., 2017). This provides a workflow to not only
quantify *ACD*, but also propagate uncertainty in *ACD* estimates arising from both field
measurement errors and uncertainty in allometric models. The first step is to estimate the
aboveground biomass (*AGB*, in kg) of individual trees using Chave et al.'s (2014) pantropical
biomass equation: $AGB = 0.067 \times (D^2 \times H \times WD)^{0.976}$. For trees with no height measurement
in the field, *H* was estimated using a locally calibrated *H–D* allometric equations, while wood
density (*WD*, in g cm$^{-3}$) values were obtained from the global wood density database (Chave et
al., 2009; Zanne et al., 2009; see Appendix S1 for additional details on both *H* and *WD*
estimation).

[13]

In addition to quantifying *AGB*, Réjou-Méchain et al.'s (2017) workflow uses Monte Carlo
simulations to propagate uncertainty in biomass estimates due to (i) measurement errors in *D*
[following Chave et al.'s (2004) approach, where 95% of stems are assumed to contain small
measurement errors that are in proportion to *D*, while the remaining 5% is assigned a gross
measurement error of 4.6 cm], (ii) uncertainty in *H–D* allometries, (iii) uncertainty in *WD*
estimates arising from incomplete taxonomic identification and/or coverage of the global wood
density database, and (iv) uncertainty in the *AGB* equation itself. Using this approach, we
generated 100 estimates of *AGB* for each recorded tree. *ACD* was then quantified by summing
the *AGB* of all trees within a plot, dividing the total by the area of the plot, and applying a
carbon content conversion factor of 0.47 (Martin and Thomas, 2011). By repeating this across
all simulated values of *AGB*, we obtained 100 estimates of *ACD* for each of the 173 plots that
reflect the uncertainty in stand-level carbon stocks. As a last step, we used data from 45 plots
in Danum Valley – where all stems with $D \geq 1$ cm were measured – to develop a correction
factor that compensates for the carbon stocks of stems with $D < 10$ cm that were not recorded
[Phillips et al., 1998; see Eq. (S2) in Appendix S1].

### 2.3.1 Stand basal area and wood density estimation

In addition to estimating *ACD* for each plot, we also calculated basal area (*BA*, in m$^2$ ha$^{-1}$) and
the community-weighted mean *WD*, as well as their uncertainties. *BA* was quantified by
summing $\pi \times (D/2)^2$ across all stems within a plot, and then applying a correction factor that
accounts for stems with $D < 10$ cm that were not measured [see Eq. (S3) in Appendix S1]. In
the case of *BA*, uncertainty arises from measurement errors in *D*, which were propagated
through following the approach of Chave et al. (2004) described in Sect. 2.3. The community-weighted mean *WD* of each plot was quantified as $\sum BA_{ij} \times WD_i$, where $BA_{ij}$ is the relative basal area of species *i* in plot *j*, and $WD_i$ is the mean wood density of species *i*. Uncertainty in plot-level *WD* reflects incomplete taxonomic information and/or lack of coverage in the global wood density database.

[14]

## 2.4    Airborne laser scanning data

ALS data covering the permanent forest plots described in Sect. 2.2 were acquired through two independent surveys, the first undertaken by NERC's Airborne Research Facility (ARF) in November of 2014 and the second by the Carnegie Airborne Observatory (CAO) in April of 2016. Table 1 specifies which plots where flown with which system. NERC ARF operated a Leica ALS50-II LiDAR sensor flown on a Dornier 228-201 at an elevation of 1400–2400 m.a.s.l. (depending on the study site) and a flight speed of 120–140 knots. The sensor emits pulses at a frequency of 120 kHz, has a field of view of 12° and a footprint of about 40 cm. The average point density was 7.3 points m$^{-2}$. The Leica ALS50-II LiDAR sensor records both discrete point and full waveform ALS, but for the purposes of this study only the discrete return data, with up to four returns recorded per pulse, were used. Accurate georeferencing of the ALS point cloud was ensured by incorporating data from a Leica base station running in the study area concurrently to the flight. The ALS data were pre-processed by NERC's Data Analysis Node and delivered in LAS format. All further processing was undertaken using LAStools software (http://rapidlasso.com/lastools). The CAO campaign was conducted using the CAO–3 system, a detailed description of which can be found in Asner et al. (2012). Briefly, CAO–3 is a custom-designed, dual-laser full-waveform system that was operated in discrete return collection mode for this project. The aircraft was flown at 3600 m.a.s.l. at a flight speed of 120–140 knots. The ALS system was set to a field of view of 34° (after 2° cut-off from each edge) and a combined-channel pulse frequency of 200 kHz. The ALS pulse footprint at 3600 m.a.s.l. was approximately 1.8 m. With adjacent flight-line overlap, these settings yielded approximately 2.0 points $m^{-2}$.

### 2.4.1  Airborne laser scanning metrics

ALS point clouds derived from both surveys were classified into ground and non-ground points, and a digital elevation model (DEM) was fitted to the ground returns to produce a raster at 1 m resolution. The DEM was then subtracted from the elevations of all non-ground returns to produce a normalised point cloud, from which a canopy height model (CHM) was constructed by averaging the first returns. Finally, any gaps in the raster of the CHM were filled by averaging neighbouring cells. From the CHMs we calculated two metrics for each of the permanent field plots: top-of-canopy height (*TCH*, in m) and canopy cover at 20 m aboveground (*Cover$_{20}$*). *TCH* is the mean height of the pixels which make up the surface of the CHM. Canopy cover is defined as the proportion of area occupied by crowns at a given height aboveground (i.e., 1 – gap fraction). *Cover$_{20}$* was calculated by creating a plane horizontal to the ground in the CHM at a height of 20 m aboveground, counting the number of pixels for which the CHM lies above the plane, and then dividing this number by the total number of pixels in the plot. A height of 20 m aboveground was chosen as previous work showed this to be the optimal height for estimating plot-level *BA* (Coomes et al., 2017).

### 2.4.2  Accounting for geopositional uncertainty

Plot coordinates obtained using a GPS are inevitably associated with a certain degree of error, particularly when working under dense forest canopies. However, this source of uncertainty is generally overlooked when attempting to relate field-estimates of *ACD* to ALS metrics. To
account for geopositional uncertainty, we introduced normally-distributed random errors in the
plot coordinates. These errors were assumed to be proportional to the operational accuracy of
the GPS unit used to geolocate a given plot: ±2 m for plots recorded with the Geneq SXBlue
II GPS and ±10 m for those geolocated using either the Garmin GPSMap60 or Garmin
GPSMAP 64S devices. This process was iterated 100 times, and at each step we calculated
*TCH* and *Cover₂₀* across all plots. Note that for plots from the SAFE project and those situated
along riparian buffer zones, ALS metrics were calculated for each individual 0.0625 ha plot
before being aggregated into blocks (as was done for the field data).

## 2.5    Modelling aboveground carbon density and associated uncertainty

We started by using data from the 173 field plots to fit a regional form of Asner & Mascaro's
(2014) model, where *ACD* is expressed as the following function of ALS-derived *TCH* and
field-based estimates of *BA* and *WD*:

$$ACD = \rho_0 \times TCH^{\rho_1} \times BA^{\rho_2} \times WD^{\rho_3} \qquad (2)$$

where $\rho_{0-3}$ represent constants to be estimated from empirical data. In order to apply Eq. (2)
to areas where field data are not available, the next step is to develop sub-models to estimate
*BA* and *WD* from ALS metrics. Of particular importance in this regard is the accurate and
unbiased estimation of *BA*, which correlates very strongly with *ACD* (Pearson's correlation
coefficient = 0.93 across the 173 plots). Asner & Mascaro (2014) found that a single ALS
metric – *TCH* – could be used to reliably estimate both *BA* and *WD* across a range of tropical
forest regions. However, recent work suggests this may not always be the case (Duncanson et al., 2015; Spriggs, 2015). In particular, Coomes et al. (2017) showed that ALS metrics that capture information about canopy cover at a given height aboveground – such as $Cover_{20}$ – were better suited to estimating $BA$ [also see Bouvier et al. (2015)]. Here we compared these two approaches to test whether $Cover_{20}$ can prove a useful metric to distinguish between forests with similar $TCH$ but substantially different $BA$.

[17]

### 2.5.1 Basal area sub-models

Asner & Mascaro (2014) modelled $BA$ as the following function of $TCH$:

$$BA = \rho_0 \times TCH \tag{3}$$

We compared the goodness of fit of Eq. (3) to a model that additionally incorporates $Cover_{20}$ as a predictor of $BA$. Doing so, however, requires accounting for the fact that $TCH$ and $Cover_{20}$ are correlated. To avoid issues of collinearity (Dormann et al., 2013), we therefore first modelled the relationship between $Cover_{20}$ and $TCH$ using logistic regression, and used the residuals of this model to identify plots that have higher or lower than expected $Cover_{20}$ for a given $TCH$:

$$\ln\left(\frac{Cover_{20}}{1 - Cover_{20}}\right) = \rho_0 + \rho_1 \times \ln(TCH) \tag{4}$$

Predicted values of canopy cover ($\widehat{Cover}_{20}$) can be obtained from Eq. (4) as follows:

$$\widehat{Cover}_{20} = \frac{1}{1 + e^{-\rho_0} \times TCH^{-\rho_1}} \tag{5}$$

From this, we calculated the residual cover ($Cover_{resid}$) for each of the 173 field plots as
$Cover_{20} - \widehat{Cover}_{20}$, and then modelled *BA* as the following non-linear function of *TCH* and
$Cover_{resid}$ :

$$BA = \rho_0 \times TCH^{\rho_1} \times (1 + \rho_2 \times Cover_{resid}) \qquad (6)$$

Eq. (6) was chosen after careful comparison with alternative functional forms. This included
modelling *BA* directly as a function of $Cover_{20}$, without including *TCH* in the regression. We
discarded this last option as *BA* estimates were found to be highly sensitive to small variations
in canopy cover when $Cover_{20}$ approaches 1.

### 2.5.2 Wood density sub-models

Following Asner & Mascaro (2014), we modelled *WD* as a power-law function of *TCH*:

$$WD = \rho_0 \times TCH^{\rho_1} \qquad (7)$$

The expectation is that, because the proportion of densely-wooded species tends to increase
during forest succession (Slik et al., 2008), taller forests should – on average – have higher
stand-level *WD* values. While this explicitly ignores the well-known fact that *WD* is also
influenced by environmental factors that have nothing to do with disturbance (e.g., soils or
climate; Quesada et al., 2012), we chose to fit a single function for all sites as from an
operational standpoint applying forest type-specific equations would require information on
the spatial distribution of these forest types across the landscape (something which may not
necessarily be available, particularly for the tropics). For comparison, we also tested whether
replacing *TCH* with $Cover_{20}$ would improve the fit of the *WD* model.

[19]

### 2.5.3   Error propagation and model validation

We developed the following approach based on leave-one-out cross validation to assess model performance and propagate uncertainty in *ACD* estimates: (i) of the 173 field plots, one was set aside for validation, while the rest were used to calibrate models; (ii) the calibration dataset was used to fit both the regional *ACD* model [Eq. (2)], as well as of the *BA* and *WD* sub-models [Eq. (3, 6–7)]; (iii) the fitted models were used to generate predictions of *BA*, *WD* and *ACD* for the validation plot previously set aside. In each case, Monte Carlo simulations were used to incorporate model uncertainty in the predicted values. For Eq. (4) and (6), parameter estimates were obtained using the L-BFGS-B nonlinear optimization routine implemented in Python (Morales and Nocedal, 2011). For power-law models fit to log-log transformed data [i.e., Eq. (2) and (7)], we applied the Baskerville (1972) correction factor by multiplying predicted values by $\exp(\sigma^2/2)$, where $\sigma$ is the estimated standard deviation of the residuals (also known as the residual standard error); (iv) model fitting and prediction steps (ii–iii) were repeated 100 times across all estimates of *ACD*, *BA*, *WD*, *TCH* and *Cover$_{20}$* that had previously been generated for each field plot. This allowed us to fully propagate uncertainty in *ACD* arising from field measurement errors, allometric models and geopositional errors; (v) lastly, steps (i–iv) were repeated for all 173 field plots.

Once predictions of *ACD* had been generated for all 173 plots, we assessed model performance by comparing predicted and observed *ACD* values ($ACD_{pred}$ and $ACD_{obs}$, respectively) on the basis of root mean square error [RMSE; calculated as $\sqrt{\frac{1}{N}\sum_{i=1}^{N}\left(ACD_{obs}-ACD_{pred}\right)^2}$] and relative systematic error [or bias; calculated as $\frac{1}{N}\sum_{i=1}^{N}\left(\frac{ACD_{pred}-ACD_{obs}}{ACD_{obs}}\right)\times 100$]. Additionally,

[20]

we tested how plot-level errors (calculated for each individual plot as $\frac{|ACD_{obs} - ACD_{pred}|}{ACD_{obs}} \times 100$)

varied as a function of forest carbon stocks and in relation to plot size (Réjou-Méchain et al.,

2014).

## 2.6    Comparison with satellite-derived estimates of aboveground carbon density

We compared the accuracy of *ACD* estimates obtained from ALS with those of two existing

carbon maps that cover the study area. The first of these is a carbon map of the SAFE landscape

and Maliau Basin derived from RapidEye satellite imagery (Pfeifer et al., 2016). The map has

a resolution of 25 × 25 m and makes use of textural and intensity information from four

wavebands to model forest biomass (which we converted to carbon by applying a conversion

factor of 0.47; Martin & Thomas, 2011). The second is a recently published consensus map of

pan-tropical forest carbon stocks at 1 km resolution (Avitabile et al., 2016). It makes use of

field data and high-resolution locally-calibrated carbon maps to refine estimates from existing

pan-tropical datasets obtained through satellite observations (Baccini et al., 2012; Saatchi et

al., 2011). To assess the accuracy of the two satellite products, we extracted *ACD* values from

both carbon maps for all overlapping field plots. For consistency with previous analyses, *ACD*

values for SAFE project plots and those in riparian buffer zones were extracted at the individual

plot level (i.e., 0.0625 ha scale) before being aggregated into the same blocks used for ALS-

model generation. In the case of Avitabile et al.'s (2016) map, *ACD* values from field plots falling within the same 1 km grid cell were averaged. We then compared field and satellite-derived estimates of *ACD* on the basis of RMSE and bias.

# 3    Results

The regional model of *ACD* – parameterized using field estimates of wood density and basal area and ALS estimates of canopy height – was:

$$ACD_{Regional} = 0.567 \times TCH^{0.554} \times BA^{1.081} \times WD^{0.186} \tag{8}$$

The model had an RMSE of 19.0 Mg C ha⁻¹ and a bias of 0.6% (Fig. 2a; see Appendix S3 for confidence intervals on all parameter estimates). The regional *ACD* model fit the data better than Asner & Mascaro's (2014) general model [i.e., Eq. (1) in the Sect. 1], which had an RMSE of 32.0 Mg C ha⁻¹ and tended to systematically underestimate *ACD* values (bias = –7.1%; Fig. 2b).

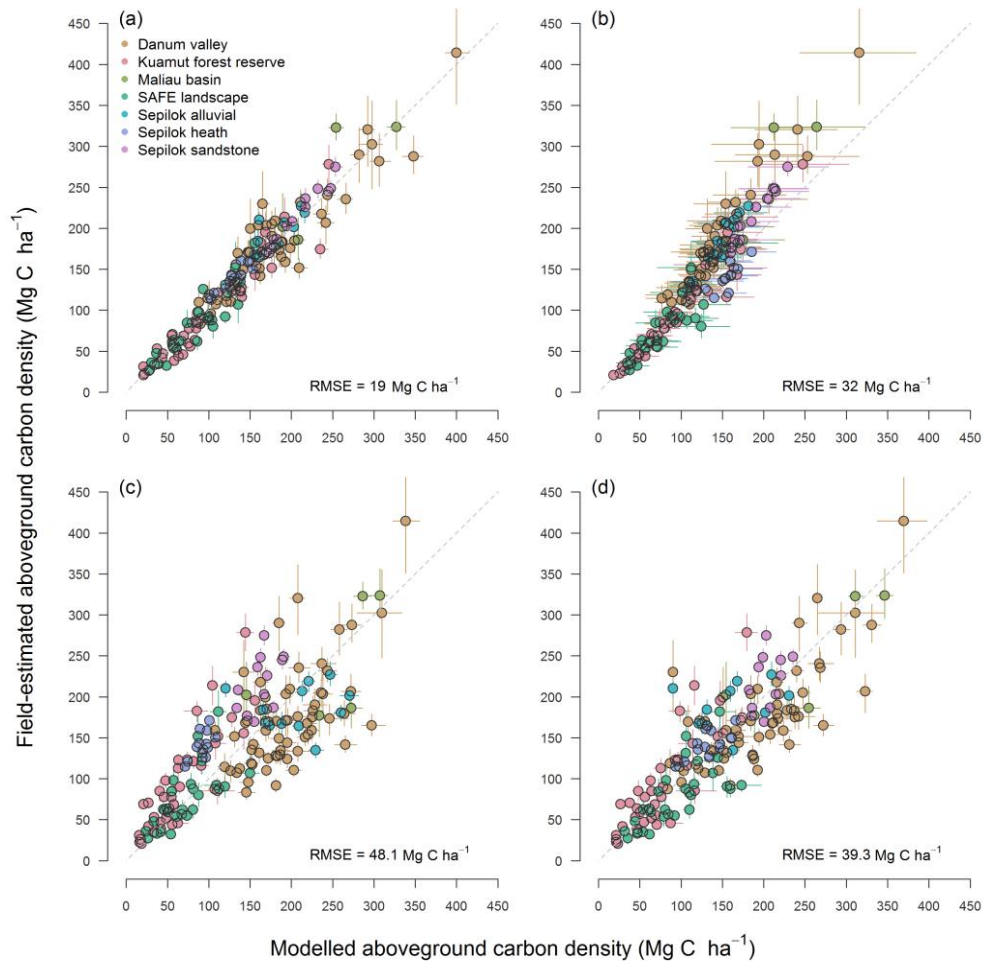

**Fig. 2:** Relationship between field-estimated and modelled aboveground carbon density (*ACD*). Panel
(**a**) shows the fit of the regionally-calibrated *ACD* model [Eq. (8) in Sect. 3] which incorporate field-
estimated basal area (*BA*) and wood density (*WD*), while (**b**) corresponds to Asner & Mascaro's (2014)
general *ACD* model [Eq. (1) in Sect. 1]. Panels (**c–d**) illustrate the predictive accuracy of the regionally-
calibrated *ACD* model when field-measured *BA* and *WD* values are replaced with estimated derived
from airborne laser scanning. In (**c**) *BA* and *WD* were estimated from top-of-canopy height (*TCH*) using
Eq. (9) and (11), respectively. In contrast, *ACD* estimates in panel (**d**) were obtained by modelling *BA*
as a function of both *TCH* and canopy cover at 20 meters aboveground following Eq. (10). In all panels,
predicted *ACD* values are based on leave-one-out cross validation. Dashed lines correspond to a 1:1

relationship. Error bars are standard deviations and the RMSE of each comparison is printed in the bottom right-hand corner of the panels.

## 3.1    Basal area sub-models

When modelling *BA* in relation to *TCH*, we found the best-fit model to be:

$$BA = 1.112 \times TCH \tag{9}$$

In comparison, when *BA* was expressed a function of both *TCH* and $Cover_{resid}$ we obtained the following model:

$$BA = 1.287 \times TCH^{0.987} \times (1 + 1.983 \times Cover_{resid}) \tag{10}$$

where $Cover_{resid} = Cover_{20} - (1 + e^{12.431} \times TCH^{-4.061})^{-1}$ (Fig. 5). Of the two sub-models used to predict *BA*, Eq. (10) proved the better fit to the data (RMSE = 9.3 and 6.6 m$^2$ ha$^{-1}$, respectively; see Appendix S2), reflecting the fact that in our case *BA* was more closely related to canopy cover than *TCH* (Fig. 3).

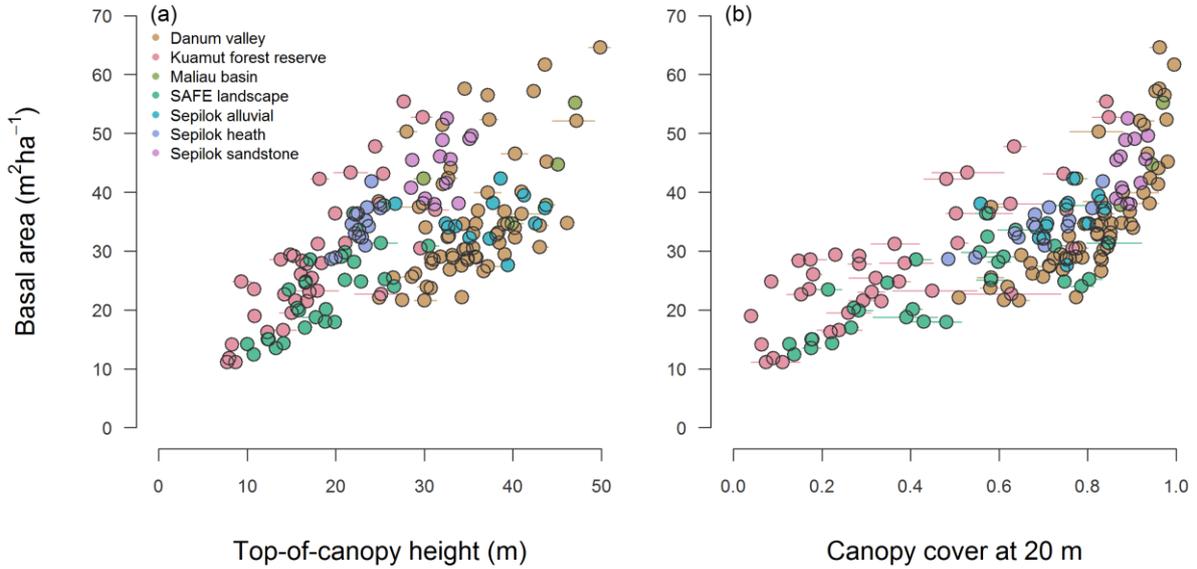

**Fig. 3:** Relationship between field-measured basal area and (**a**) top-of-canopy height and (**b**) canopy cover at 20 meters aboveground as measured through airborne laser scanning. Error bars correspond to standard deviations.

### 3.2    Wood density sub-model

When modelling *WD* as a function of *TCH*, we found the best fit model to be:

$$WD = 0.385 \times TCH^{0.097} \tag{11}$$

Across the plot network *WD* showed a general tendency to increase with *TCH* (Fig. 4; RMSE of 0.056 g cm$^{-3}$). However, the relationship was weak and Eq. (11) did not capture variation in *WD* equally well across the different forest types (see Appendix S2). In particular, heath forests at Sepilok – which have very high *WD* despite being much shorter than surrounding lowland dipterocarp forests (0.64 against 0.55 g cm$^{-3}$) – were poorly captured by the *WD* sub-model. We found no evidence to suggest that replacing *TCH* with canopy cover at 20 m aboveground would improve the accuracy of these estimates (see Appendix S2).

[25]

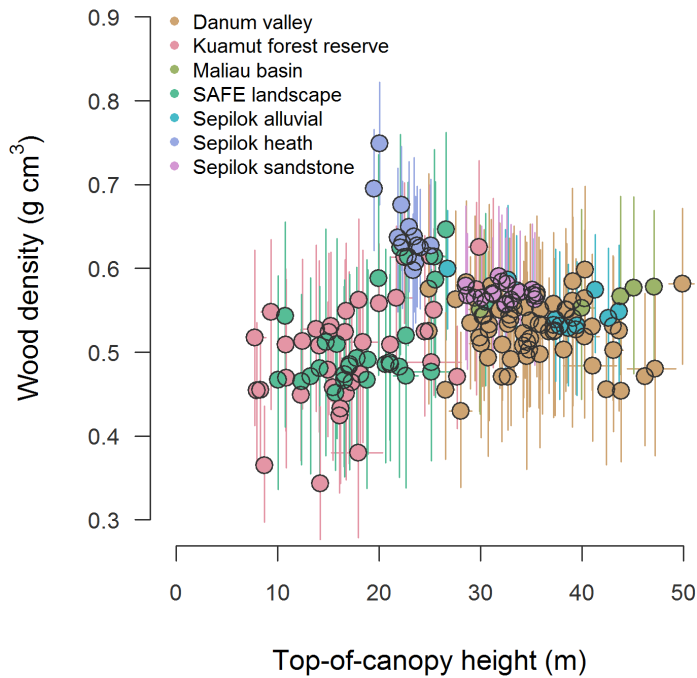

**Fig. 4:** Relationship between community-weighted mean wood density (from field measurements) and top-of-canopy height (from airborne laser scanning). Error bars correspond to standard deviations.

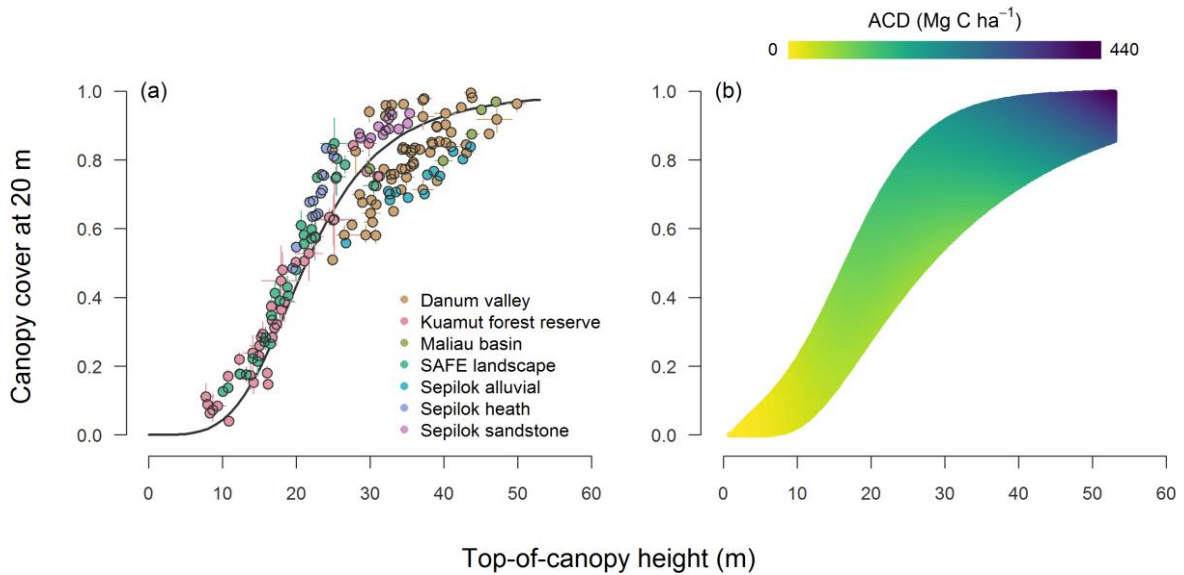

**Fig. 5:** Relationship between ALS-derived canopy cover at 20 meters aboveground and top-of-canopy height. Panel (**a**) shows the distribution of the field plots with a line of best fit passing through the data. Panel (**b**) illustrates how estimates of aboveground carbon density [*ACD*; obtained using Eq. (8), with Eq. (10) and (11) as inputs] vary as a function of the two ALS metrics for the range of values observed across the forests of Sabah.

## 3.3    Estimating aboveground carbon density from airborne laser scanning

When field-based estimates of *BA* and *WD* were replaced with ones derived from *TCH* using Eq. (9) and (11), the regional *ACD* model generated unbiased estimates of *ACD* (bias = −1.8%). However, the accuracy of the model decreased substantially (RMSE = 48.1 Mg C ha$^{-1}$; Fig. 2c). In particular, the average plot-level error was 21% and remained relatively constant across the range of *ACD* values observed in the field data (Fig. 6a). In contrast, when the combination of *TCH* and *Cover$_{20}$* were used to estimate *BA* through Eq. (10), we obtained more accurate estimates of *ACD* (RMSE = 39.3 Mg C ha$^{-1}$, bias = 5.3%; Fig. 2d). Moreover, in this instance plot-level errors showed a clear tendency to decrease in large and high-carbon density plots (Fig. 6).

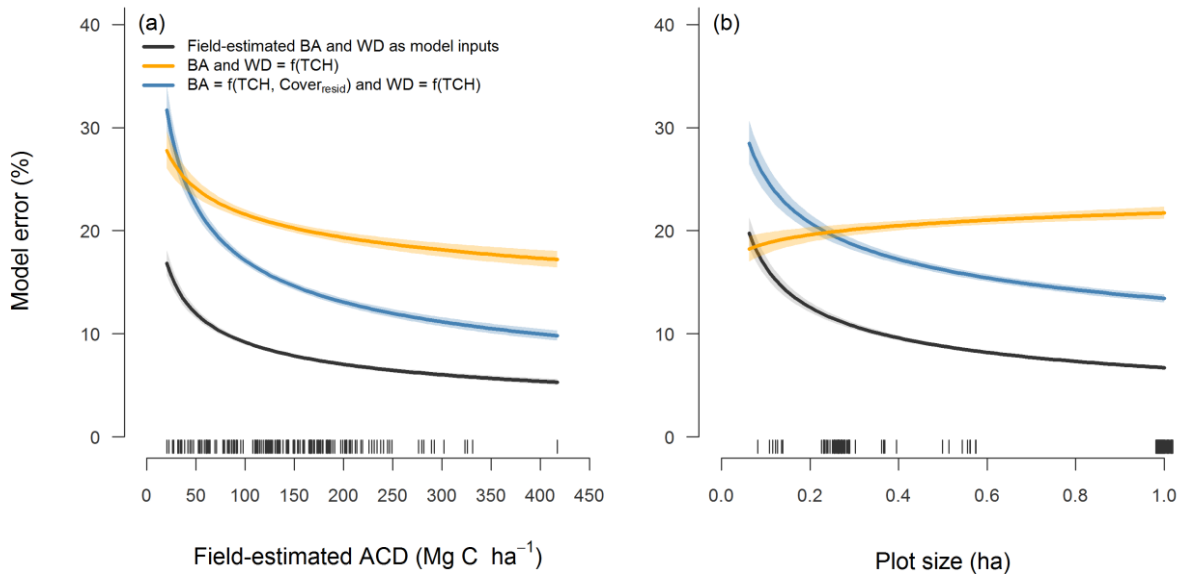

**Fig. 6:** Model errors [calculated for each individual plot as $\left(\left|ACD_{obs} - ACD_{pred}\right|\right)/ACD_{obs} \times 100$] in relation to (**a**) field-estimated aboveground carbon density (*ACD*) and (**b**) plot size. Curves (± 95% shaded confidence intervals) were obtained by fitting linear models to log-log transformed data. Black lines correspond to the regionally-calibrated *ACD* model [Eq. (8) in Sect. 3]. Orange lines show model errors when basal area (*BA*) was estimated from top-of-canopy height (*TCH*) using Eq. (9). In contrast, blue lines show model errors when *BA* was expressed as a function of both *TCH* and canopy cover at 20 meters aboveground following Eq. (10). Vertical dashed lines along the horizontal axis show the distribution of the data [in panel (**b**) plot size values were jittered to avoid overlapping lines].

## 3.4    Comparison with satellite-derived estimates of aboveground carbon density

When compared to ALS-derived estimates of *ACD*, both satellite-based carbon maps of the study area showed much poorer agreement with field data (Fig. 7). Pfeifer et al.'s (2016) map covering the SAFE landscape and Maliau Basin systematically underestimated *ACD* (bias = – 36.9%) and had an RMSE of 77.8 Mg C ha[-1]. In contrast, Avitabile et al.'s (2016) pan-tropical map overestimated carbon stocks by 111.2% on average, and had an RMSE of 100.1 Mg C ha[-1].

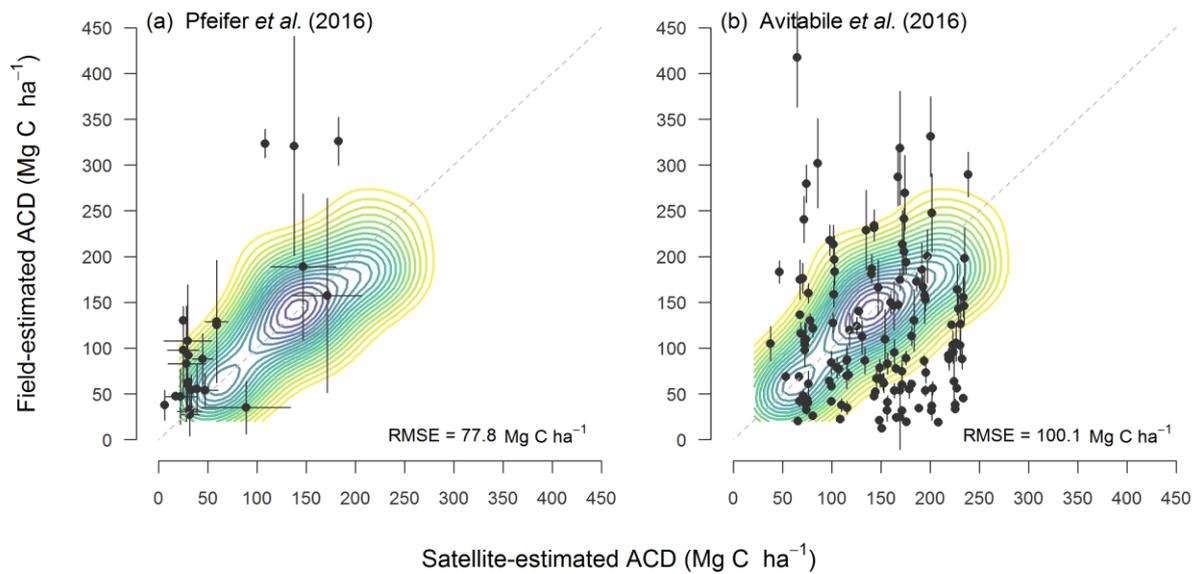

**Fig. 7:** Comparison between field-estimated aboveground carbon density (*ACD*) and satellite-derived estimates of *ACD* reported in (**a**) Pfeifer et al. (2016) and (**b**) Avitabile et al. (2016). Error bars correspond to standard deviations, while the RMSE of the satellite estimates is printed in the bottom right-hand corner of the panels. For comparison with *ACD* estimates obtained from airborne laser scanning, a kernel density plot fit to the points in Fig. 2d is displayed in the background.

## 4       Discussion

We developed an area-based model for estimating aboveground carbon stock from ALS data that can be applied to mapping the lowland tropical forests of Borneo. We found that adding a canopy cover term to Asner & Mascaro's (2014) general model substantially improved its goodness of fit (Fig. 2c–d), as it allowed us to capture variation in stand basal area much more effectively compared to models parameterized solely using plot-averaged *TCH*. In this process, we also implemented an error propagation approach that allows various sources of uncertainty in *ACD* estimates to be incorporated into carbon mapping efforts. In the following sections we place our approach in the context of ongoing efforts to use remotely sensed data to monitor forest carbon stocks, starting with ALS-based approaches and then comparing these to satellite-based modelling. Finally, we end by discussing the implication of this work for the conservation of Borneo's forests.

### 4.1       Including canopy cover in the Asner & Mascaro (2014) carbon model

We find that incorporating a measure of canopy cover at 20 m aboveground in the Asner & Mascaro (2014) model improves its goodness-of-fit substantially without compromising its generality. Asner & Mascaro's (2014) model is grounded in forest and tree geometry, drawing its basis from allometric equations for estimating tree aboveground biomass such as that of Chave et al. (2014), where a tree's biomass is expressed as a multiplicative function of its diameter, height and wood density: $AGB = \rho_0 \times (WD \times D^2 \times H)^{\rho_1}$. By analogy, the carbon stock within a plot is related to the product of mean wood density, total basal area and top-of-canopy height (each raised to a power). Deriving this power-law function from a knowledge of the tree size distribution and tree-biomass relationship is far from straightforward mathematically

[30]

(Spriggs, 2015; Vincent et al., 2014), but this analogy seems to hold up well in a practical sense. When fitted to data from 14 forest types spanning aridity gradients in the Neotropics and Madagascar, Asner & Mascaro (2014) found that a single relationship applied to all forests types, once regional differences in structure were incorporated as sub-models relating *BA* and *WD* to *TCH*. However, the model's fit depends critically on there being a close relationship between *BA* and *TCH*, as *BA* and *ACD* tend to be tightly coupled. Whilst that held true for the 14 forest types previously studied, in Bornean forest we found that the *BA* sub-models could be improved considerably by including canopy cover as an explanatory variable. This makes intuitive sense if one considers an open forest comprised of just a few trees – the crown area of each tree scales with its basal area, so the gap fraction at ground level of a plot is negatively related to the basal area of its trees (Singh et al., 2016). A similar principle applies in denser forests, but in forests with multiple tiers formed by overlapping canopies such as those that occur in Borneo, the best-fitting relationship between gap fraction and basal area is no longer at ground level, but is instead further up the canopy (Coomes et al., 2017). We refer to the refined model as the generalised geometric scaling model.

The functional form used to model *BA* in relation to *TCH* and residual forest cover [i.e., Eq. (10)] was selected for two reasons: first, for a plot with average canopy cover, the model reduces to the classic model of Asner & Mascaro (2014), making comparisons straightforward. Secondly, simpler functional forms (e.g., ones relating *BA* directly to *Cover*$_{20}$) were found to have very similar goodness-of-fit, but predicted unrealistically high *ACD* estimates for a small fraction of pixels when applied to mapping carbon across the landscape. This study is the first to formally introduce canopy cover into the modelling framework of Asner & Mascaro (2014), but several other studies have concluded that gap fraction is an important variable to include in
multiple regression models of forest biomass (Colgan et al., 2012; Singh et al., 2016; Spriggs,
2015). Regional calibration of the Asner & Mascaro (2014) model was necessary for the
lowland forests of Southeast Asia [also see Coomes et al. (2017)], because dominance by
dipterocarp species make them structurally distinct (Ghazoul, 2016): trees in the region grow
tall but have narrow stems for their height (Banin et al., 2012; Feldpausch et al., 2011), creating
forests that have among the greatest carbon densities of any in the tropics (Avitabile et al.,
2016; Sullivan et al., 2017).

Our approach differs from the multiple-regression-with-model-selection approach that is
typically adopted for modelling *ACD* of tropical forests using ALS data (Chen et al., 2015;
Clark et al., 2011; D'Oliveira et al., 2012; Drake et al., 2002; Hansen et al., 2015; Ioki et
al., 2014; Jubanski et al., 2013; Réjou-Méchain et al., 2015; Singh et al., 2016). These
studies – which build on two decades of research in temperate and boreal forests (Lefsky et al.,
1999; Nelson et al., 1988; Popescu et al., 2011; Wulder et al., 2012) – typically calculate
between 5 and 25 summary statistics from the height distribution of ALS returns and explore
the performance of models constructed using various combinations of those summary
statistics as explanatory variables. Typically, the "best-supported" model is then selected
from the list of competing models on offer, by comparing relative performance using
evaluation statistics such as $R^2$, RMSE or AIC.

There is no doubt that selecting regression models in this way provides a solid basis for making
model-assisted inferences about regional carbon stocks and their uncertainty (Ene et al.,
2012; Gregoire et al., 2016). However, a well-recognised problem is that models tend to be idiosyncratic by virtue of local fine-tuning, so cannot be applied more widely than the region for which they were calibrated, and cannot be compared very easily with other studies. For example, it comes as no surprise that almost all publications identify mean height or some metric of upper-canopy height (e.g., 90th or 99th percentile of the height distribution) as being the strongest determinant of biomass. But different choices of height metric make these models difficult to compare. Other studies include variance terms to improve goodness of fit. For instance, a combination of 75th quantile and variance of return heights proved effective in modelling *ACD* of selectively logged forests in Brazil (D'Oliveira et al., 2012). Several recent studies include measures of laser penetration to the lower canopy in the best-performing models. A model developed for lower montane forests in Sabah included the proportion of last returns within 12 m of the ground (Ioki et al., 2014), while the proportions of returns in various height tiers were selected for ALS carbon mapping of sub-montane forest in Tanzania (Hansen et al., 2015). Working with Asner & Mascaro's (2014) power-law model may sacrifice goodness-of-fit compared with locally tuned multiple regression models, but provides a systematic framework for ALS modelling of forest carbon throughout the tropics.

## 4.2 Quantifying and propagating uncertainty

One of the most important applications of *ACD*-estimation models is to infer carbon stocks within regions of interest (e.g., REDD project, state or country). Carbon stock estimation has traditionally been achieved by networks of inventory plots, designed to provide unbiased estimates of timber volumes within an acceptable level of uncertainty, using well-established design-based approaches (Särndal et al., 1992). Forest inventories are increasingly supported by the collection of cost-effective auxiliary variables, such as ALS-estimated forest height and cover, that increase the precision of carbon stock estimation when used to construct regression models, which are in turn used to estimate carbon in an area where the auxiliary variables have been measured (e.g., McRoberts et al., 2013). Assessment of uncertainty within this model-assisted framework requires uncertainty to be quantified and propagation through all process involved in the calculation of landscape carbon stocks and statistical models (Chen et al., 2015). Our study propagates errors related to field measurements and allometric models using the framework developed in the recently published *BIOMASS* package (Réjou-Méchain et al., 2017), and those related to ALS data and *ACD* models with custom-made code to estimate plot-level uncertainty (Chen et al., 2015). This approach, which is fundamentally different to estimating uncertainty by comparing model predictions to validation field plots, is rarely used in remote sensing (e.g., Gonzalez et al., 2010), but is the more appropriate technique to use when there is uncertainty in field measurements (Chen et al., 2015). Our Monte Carlo framework allows field-measurement errors, geo-positional errors and model uncertainty to be propagated in a straightforward manner (Yanai et al., 2010). An additional step would be to also incorporate spatial autocorrelation in the model (McRoberts et al., 2016), something which we attempted but found to increase the bias in model predictions (Appendix S4; see also Rejou-Mechain et al. (2014). *ACD* uncertainty of model which leverages both TCH and Cover$_{20}$ to estimate carbon stock decreases strongly with increasing plot size and forest biomass, while the trend is much less marked with the *TCH*-only model (Fig. 6), signalling another advantage of this approach.

Several sources of potential bias remain. Community-weighted wood density is only weakly related to ALS metrics and is estimated with large errors (Fig. 4). The fact that wood density cannot be measured remotely is well recognised, and the assumptions used to map wood density from limited field data have major implications for carbon maps produced by satellites (Mitchard et al., 2014). For Borneo, it may prove necessary to develop separate wood density sub-models for estimating carbon in heath forests versus other lowland forest types. Height allometry is another source of potential bias (Rutishauser et al., 2013): four published height-diameter curves for Sabah show similar fits for small trees (<50 cm diameter) but diverge for large trees, which contain most of the biomass (Coomes et al., 2017). Terrestrial laser scanning is likely to address this issue in the coming years (Calders et al., 2015). Another issue is that ALS point cloud densities affect *TCH* estimates, particularly at low point densities (Gobakken and Næsset, 2008; Manuri et al., 2014), so care is needed when applying the model to different surveys [although we note that *TCH* is less sensitive to flight specifications than other ALS structural metrics; Asner & Mascaro (2014)]. Finally, our model estimates aboveground carbon stocks, but much of the forest's carbon is held in other pools: for example, in a selectively logged dipterocarp forest in Sabah, Saner et al. (2012) recorded carbon stocks of 168 Mg C ha$^{-1}$ partitioned into aboveground carbon in trees (55%), belowground carbon in trees (10%), deadwood (8%) and soil organic matter (24%), understory vegetation (3%), and standing litter (<1%). Almost half that carbon is not being estimated by using ALS data.

## 4.3    Comparison with satellite-derived maps

Our results show that when compared to independent field data, existing satellite products systematically under- or over-estimate *ACD* (depending on the product; Fig. 7). Whilst the findings need to be treated with caution, because field plots are smaller than pixel sizes [see Rejou-Mechain et al. (2014)], it does appear that ALS provides much more robust and accurate estimates of *ACD* and its heterogeneity within the landscape than possible with current space-borne sensors. However, ALS data is limited in its temporal and spatial coverage, due to high operational costs. Consequently, researchers should focus on fusing ALS-derived maps of *ACD* with satellite data to get wall-to-wall coverage in space and time. NASA plans to start making high resolution laser ranging observations from the international space station in 2018 as part of the GEDI mission, while ESA's biomass mission will use P-band synthetic aperture radar to monitor forests from space from 2020. Pan-tropical monitoring of forest carbon using data from a combination of space-borne sensors is fast approaching, and regional carbon equations derived from ALS data such as the one we develop here will be critical to calibrate and validate these efforts.

## 5        Conclusions: applications of carbon-estimation models for forest conservation

This paper has developed a model for estimating carbon stock from ALS data that is applicable to lowland dipterocarp forests of Borneo and Southeast Asia more widely. The model is grounded in the fundamentals of tree geometry, but emphasizes the importance of variance in height (i.e., canopy cover) alongside forest height [see Bouvier et al. (2015)]. With the development of plans to reduce forest losses and/or recover forests from previous logging- and fire-driven degradation, methods will be needed to support spatially-explicit monitoring of changing forest carbon stocks over time (Putz et al., 2012). Small-footprint ALS will be an important contributor to this effort. The current policy environment within Southeast Asia emphasises demand for the development of rapid aerial monitoring approaches that are applicable at landscape and regional scales. Forest loss across the region, and especially in Borneo, has been driven by expansion of large-scale industrial palm oil plantations since the 1970s, and much of this planting has replaced carbon-dense forests (Carlson et al., 2012a, 2012b; Gaveau et al., 2014). Certification bodies such as the Roundtable on Sustainable Palm Oil (RSPO) have responded to criticisms by adopting policies that prohibit planting on land designated as High Conservation Value (HCV), and prescribe compensation mechanisms such as restoration tree planting when these principles are violated by their members (RSPO, 2007). More recently, RSPO have proposed supplementing the HCV approach with High Carbon Stock (HCS) assessments that would restrict expansion of palm oil plantations onto carbon-dense forests.

The methods proposed for identifying HCS forests currently emphasise the use of satellite imagery for defining vegetation classes at coarse resolution, followed by the establishment of permanent plot networks to estimate carbon stocks within vegetation classes from field measurements (HCS Approach Steering Group, 2015). ALS has the potential to enhance the scale and precision of these carbon stock assessments, and features much more prominently in revised methods currently in preparation (HCS Approach, 2016). The ultimate aim of these "no net deforestation" approaches is to encourage agricultural development on land that has already been cleared of HCS forests, which includes widespread anthropogenic tropical grasslands (Corley, 2009). The extent to which identifying and protecting HCS forests will have significant co-benefits for conservation depends on the spatial congruence of biodiversity and carbon, which is contentious and context-specific (Beaudrot et al., 2016; Sullivan et al., 2017). Identifying these potential co-benefits becomes all the more challenging in the face of projected climate change, which exacerbates the difficulty of spatial planning for maximising for both biodiversity and carbon storage (Struebig et al., 2015). Much less contentious, however, is that restoring degraded and heavily logged tropical forest lands through tree planting and other

silvicultural techniques, such as climber cutting, can deliver substantial benefits for biomass

and carbon storage (Hector et al., 2011), especially when conducted with a careful selection of

planting materials (Kettle et al., 2012). About 500 million ha of tropical lands are degraded

(Lamb et al., 2005), and ALS techniques will be ideally-suited to quantifying the spatial and temporal dynamics of carbon stock recovery once work to restore tropical forests scales up.

## Acknowledgements

This publication is a contribution from the NERC-funded Biodiversity And Land-use Impacts
on Tropical Ecosystem Function (BALI) consortium (http://bali.hmtf.info; grant number:
NE/K016377/1). We are grateful to NERC's Airborne Research Facility and Data Analysis
Node for conducting the survey and pre-processing the airborne data, and to Abdullah Ghani
for manning the GPS base station. D.A. Coomes was supported by an International Academic
Fellowship from the Leverhulme Trust. The Carnegie Airborne Observatory portion of the
study was supported by the UN Development Programme, Avatar Alliance Foundation,
Roundtable on Sustainable Palm Oil, World Wildlife Fund, and the Rainforest Trust. The
Carnegie Airborne Observatory is made possible by grants and donations to G.P. Asner from
the Avatar Alliance Foundation, Margaret A. Cargill Foundation, David and Lucile Packard
Foundation, Gordon and Betty Moore Foundation, Grantham Foundation for the Protection of
the Environment, W. M. Keck Foundation, John D. and Catherine T. MacArthur Foundation,
Andrew Mellon Foundation, Mary Anne Nyburg Baker and G. Leonard Baker Jr, and William
R. Hearst III. The SAFE project was supported by Sime Darby Foundation. We acknowledge
the SAFE management team, Maliau Basin Management Committee, Danum Valley
Management Committee, South East Asia Rainforest Research Partnership, Sabah Foundation,
Benta Wawasan, the State Secretary, Sabah Chief Minister's Departments, Sabah Forestry
Department, Sabah Biodiversity Centre and the Economic Planning Unit for their support,
access to the field sites and for permission to carry out fieldwork in Sabah. M. Svátek was
funded through a grant from the Ministry of Education, Youth and Sports of the Czech
Republic (grant number: INGO II LG15051) and J. Kvasnica was funded through an IGA grant

(grant number: LDF_VP_2015038). We are grateful to the many field assistants who contributed to data collection.

# References

Agrawal, A., Nepstad, D., Chhatre, A., 2011. Reducing emissions from deforestation and forest
degradation. Annu. Rev. Environ. Resour. 36, 373–396. doi:10.1146/annurev-environ-
042009-094508

Anderson-Teixeira, K.J., Davies, S.J., Bennett, A.C., Gonzalez-Akre, E.B., Muller-Landau,
H.C., Joseph Wright, S., Abu Salim, K., Almeyda Zambrano, A.M., Alonso, A., Baltzer,
J.L., Basset, Y., Bourg, N.A., Broadbent, E.N., Brockelman, W.Y., Bunyavejchewin, S.,
Burslem, D.F.R.P., Butt, N., Cao, M., Cardenas, D., Chuyong, G.B., Clay, K., Cordell, S.,
Dattaraja, H.S., Deng, X., Detto, M., Du, X., Duque, A., Erikson, D.L., Ewango, C.E.N.,
Fischer, G.A., Fletcher, C., Foster, R.B., Giardina, C.P., Gilbert, G.S., Gunatilleke, N.,
Gunatilleke, S., Hao, Z., Hargrove, W.W., Hart, T.B., Hau, B.C.H., He, F., Hoffman,
F.M., Howe, R.W., Hubbell, S.P., Inman-Narahari, F.M., Jansen, P.A., Jiang, M.,
Johnson, D.J., Kanzaki, M., Kassim, A.R., Kenfack, D., Kibet, S., Kinnaird, M.F., Korte,
L., Kral, K., Kumar, J., Larson, A.J., Li, Y., Li, X., Liu, S., Lum, S.K.Y., Lutz, J.A., Ma,
K., Maddalena, D.M., Makana, J.-R., Malhi, Y., Marthews, T., Mat Serudin, R.,
McMahon, S.M., McShea, W.J., Memiaghe, H.R., Mi, X., Mizuno, T., Morecroft, M.,
Myers, J.A., Novotny, V., de Oliveira, A.A., Ong, P.S., Orwig, D.A., Ostertag, R., den
Ouden, J., Parker, G.G., Phillips, R.P., Sack, L., Sainge, M.N., Sang, W., Sri-ngernyuang,
K., Sukumar, R., Sun, I.-F., Sungpalee, W., Suresh, H.S., Tan, S., Thomas, S.C., Thomas,
D.W., Thompson, J., Turner, B.L., Uriarte, M., Valencia, R., Vallejo, M.I., Vicentini, A.,
Vrška, T., Wang, X., Wang, X., Weiblen, G., Wolf, A., Xu, H., Yap, S., Zimmerman, J.,
2015. CTFS-ForestGEO: a worldwide network monitoring forests in an era of global

change. Glob. Chang. Biol. 21, 528–549. doi:10.1111/gcb.12712

Asner, G.P., Knapp, D.E., Boardman, J., Green, R.O., Kennedy-Bowdoin, T., Eastwood, M.,
Martin, R.E., Anderson, C., Field, C.B., 2012. Carnegie Airborne Observatory-2:
increasing science data dimensionality via high-fidelity multi-sensor fusion. Remote Sens.
Environ. 124, 454–465. doi:10.1016/j.rse.2012.06.012

Asner, G.P., Knapp, D.E., Martin, R.E., Tupayachi, R., Anderson, C.B., Mascaro, J., Sinca, F.,
Chadwick, K.D., Higgins, M., Farfan, W., Llactayo, W., Silman, M.R., 2014. Targeted
carbon conservation at national scales with high-resolution monitoring. Proc. Natl. Acad.
Sci. 111, E5016–E5022. doi:10.1073/pnas.1419550111

Asner, G.P., Mascaro, J., 2014. Mapping tropical forest carbon: calibrating plot estimates to a
simple LiDAR metric. Remote Sens. Environ. 140, 614–624.
doi:10.1016/j.rse.2013.09.023

Asner, G.P., Powell, G.V.N., Mascaro, J., Knapp, D.E., Clark, J.K., Jacobson, J., Kennedy-
Bowdoin, T., Balaji, A., Paez-Acosta, G., Victoria, E., Secada, L., Valqui, M., Hughes,
R.F., 2010. High-resolution forest carbon stocks and emissions in the Amazon. Proc. Natl.
Acad. Sci. 107, 16738–16742. doi:10.1073/pnas.1004875107

Asner, G.P., Rudel, T.K., Aide, T.M., DeFries, R., Emerson, R., 2009. A contemporary
assessment of change in humid tropical forests. Conserv. Biol. 23, 1386–1395.
doi:10.1111/j.1523-1739.2009.01333.x

Avitabile, V., Herold, M., Heuvelink, G.B.M., Lewis, S.L., Phillips, O.L., Asner, G.P.,

Armston, J., Asthon, P., Banin, L.F., Bayol, N., Berry, N., Boeckx, P., de Jong, B., DeVries, B., Girardin, C., Kearsley, E., Lindsell, J.A., Lopez-Gonzalez, G., Lucas, R., Malhi, Y., Morel, A., Mitchard, E., Nagy, L., Qie, L., Quinones, M., Ryan, C.M., Slik, F., Sunderland, T., Vaglio Laurin, G., Valentini, R., Verbeeck, H., Wijaya, A., Willcock, S., 2016. An integrated pan-tropical biomass map using multiple reference datasets. Glob. Chang. Biol. 22, 1406–1420. doi:10.1111/gcb.13139

Baccini, A., Goetz, S.J., Walker, W.S., Laporte, N.T., Sun, M., Sulla-Menashe, D., Hackler, J., Beck, P.S.A., Dubayah, R., Friedl, M.A., Samanta, S., Houghton, R.A., 2012. Estimated carbon dioxide emissions from tropical deforestation improved by carbon-density maps. Nat. Clim. Chang. 2, 182–185. doi:10.1038/nclimate1354

Banin, L., Feldpausch, T.R., Phillips, O.L., Baker, T.R., Lloyd, J., Affum-Baffoe, K., Arets, E.J.M.M., Berry, N.J., Bradford, M., Brienen, R.J.W., Davies, S., Drescher, M., Higuchi, N., Hilbert, D.W., Hladik, A., Iida, Y., Salim, K.A., Kassim, A.R., King, D.A., Lopez-Gonzalez, G., Metcalfe, D., Nilus, R., Peh, K.S.H., Reitsma, J.M., Sonké, B., Taedoumg, H., Tan, S., White, L., Wöll, H., Lewis, S.L., 2012. What controls tropical forest architecture? Testing environmental, structural and floristic drivers. Glob. Ecol. Biogeogr. 21, 1179–1190. doi:10.1111/j.1466-8238.2012.00778.x

Baraloto, C., Hardy, O.J., Paine, C.E.T., Dexter, K.G., Cruaud, C., Dunning, L.T., Gonzalez, M.A., Molino, J.F., Sabatier, D., Savolainen, V., Chave, J., 2012. Using functional traits and phylogenetic trees to examine the assembly of tropical tree communities. J. Ecol. 100, 690–701. doi:10.1111/j.1365-2745.2012.01966.x

Baskerville, G.L., 1972. Use of logarithmic regression in the estimation of plant biomass. Can.
J. For. Res. 2, 49–53.

Beaudrot, L., Kroetz, K., Alvarez-Loayza, P., Amaral, I., Breuer, T., Fletcher, C., Jansen, P.A.,
Kenfack, D., Lima, M.G.M., Marshall, A.R., Martin, E.H., Ndoundou-Hockemba, M.,
O'Brien, T., Razafimahaimodison, J.C., Romero-Saltos, H., Rovero, F., Roy, C.H., Sheil,
D., Silva, C.E.F., Spironello, W.R., Valencia, R., Zvoleff, A., Ahumada, J., Andelman,
S., 2016. Limited carbon and biodiversity co-benefits for tropical forest mammals and
birds. Ecol. Appl. 26, 1098–1111. doi:10.1890/15-0935

Bouvier, M., Durrieu, S., Fournier, R.A., Renaud, J.P., 2015. Generalizing predictive models
of forest inventory attributes using an area-based approach with airborne LiDAR data.
Remote Sens. Environ. 156, 322–334. doi:10.1016/j.rse.2014.10.004

Calders, K., Newnham, G., Burt, A., Murphy, S., Raumonen, P., Herold, M., Culvenor, D.,
Avitabile, V., Disney, M., Armston, J., Kaasalainen, M., 2015. Nondestructive estimates
of above-ground biomass using terrestrial laser scanning. Methods Ecol. Evol. 6, 198–
208. doi:10.1111/2041-210X.12301

Carlson, K.M., Curran, L.M., Asner, G.P., Pittman, A.M., Trigg, S.N., Marion Adeney, J.,
2012a. Carbon emissions from forest conversion by Kalimantan oil palm plantations. Nat.
Clim. Chang. 3, 283–287. doi:10.1038/nclimate1702

Carlson, K.M., Curran, L.M., Ratnasari, D., Pittman, A.M., Soares-Filho, B.S., Asner, G.P.,
Trigg, S.N., Gaveau, D. a, Lawrence, D., Rodrigues, H.O., 2012b. Committed carbon

emissions, deforestation, and community land conversion from oil palm plantation expansion in West Kalimantan, Indonesia. Proc. Natl. Acad. Sci. 109, 1–6. doi:10.1073/pnas.1200452109

Chave, J., Condit, R., Aguilar, S., Hernandez, A., Lao, S., Perez, R., 2004. Error propagation and scaling for tropical forest biomass estimates. Philos. Trans. R. Soc. B 359, 409–20. doi:10.1098/rstb.2003.1425

Chave, J., Coomes, D.A., Jansen, S., Lewis, S.L., Swenson, N.G., Zanne, A.E., 2009. Towards a worldwide wood economics spectrum. Ecol. Lett. 12, 351–366. doi:10.1111/j.1461-0248.2009.01285.x

Chave, J., Réjou-Méchain, M., Búrquez, A., Chidumayo, E., Colgan, M.S., Delitti, W.B.C., Duque, A., Eid, T., Fearnside, P.M., Goodman, R.C., Henry, M., Martínez-Yrízar, A., Mugasha, W.A., Muller-Landau, H.C., Mencuccini, M., Nelson, B.W., Ngomanda, A., Nogueira, E.M., Ortiz-Malavassi, E., Pélissier, R., Ploton, P., Ryan, C.M., Saldarriaga, J.G., Vieilledent, G., 2014. Improved allometric models to estimate the aboveground biomass of tropical trees. Glob. Chang. Biol. 20, 3177–3190. doi:10.1111/gcb.12629

Chen, Q., Vaglio Laurin, G., Valentini, R., 2015. Uncertainty of remotely sensed aboveground biomass over an African tropical forest: Propagating errors from trees to plots to pixels. Remote Sens. Environ. 160, 134–143. doi:10.1016/j.rse.2015.01.009

Clark, M.L., Roberts, D.A., Ewel, J.J., Clark, D.B., 2011. Estimation of tropical rain forest aboveground biomass with small-footprint lidar and hyperspectral sensors. Remote Sens.

Environ. 115, 2931–2942. doi:10.1016/j.rse.2010.08.029

Colgan, M.S., Asner, G.P., Levick, S.R., Martin, R.E., Chadwick, O.A., 2012. Topo-edaphic
controls over woody plant biomass in South African savannas. Biogeosciences 9, 1809–
1821. doi:10.5194/bg-9-1809-2012

Coomes, D.A., Dalponte, M., Jucker, T., Asner, G.P., Banin, L.F., Burslem, D.F.R.P., Lewis,
S.L., Nilus, R., Phillips, O., Phuag, M.-H., Qiee, L., 2017. Area-based vs tree-centric
approaches to mapping forest carbon in Southeast Asian forests with airborne laser
scanning data. Remote Sens. Environ. 194, 77–88. doi:10.1016/j.rse.2017.03.017

Corley, R.H.V., 2009. How much palm oil do we need? Environ. Sci. Policy 12, 134–139.
doi:10.1016/j.envsci.2008.10.011

D'Oliveira, M.V.N., Reutebuch, S.E., McGaughey, R.J., Andersen, H.-E., 2012. Estimating
forest biomass and identifying low-intensity logging areas using airborne scanning lidar
in Antimary State Forest, Acre State, Western Brazilian Amazon. Remote Sens. Environ.
124, 479–491. doi:10.1016/j.rse.2012.05.014

Dent, D.H., Bagchi, R., Robinson, D., Majalap-Lee, N., Burslem, D.F.R.P., 2006. Nutrient
fluxes via litterfall and leaf litter decomposition vary across a gradient of soil nutrient
supply in a lowland tropical rain forest. Plant Soil 288, 197–215. doi:10.1007/s11104-
006-9108-1

DeWalt, S.J.S.S.J., Ickes, K., Nilus, R., Harms, K.E.K., Burslem, D.D.F.R.P., 2006. Liana
habitat associations and community structure in a Bornean lowland tropical forest. Plant

Ecol. 186, 203–216. doi:10.1007/s11258-006-9123-6

Dormann, C.F., Elith, J., Bacher, S., Buchmann, C., Carl, G., Carré, G., Marquéz, J.R.G.,
Gruber, B., Lafourcade, B., Leitão, P.J., Münkemüller, T., McClean, C., Osborne, P.E.,
Reineking, B., Schröder, B., Skidmore, A.K., Zurell, D., Lautenbach, S., 2013.
Collinearity: A review of methods to deal with it and a simulation study evaluating their
performance. Ecography. 36, 27–46. doi:10.1111/j.1600-0587.2012.07348.x

Drake, J.B., Dubayah, R.O., Clark, D.B., Knox, R.G., Blair, J.B.B., Hofton, M.A., Chazdon,
R.L., Weishampel, J.F., Prince, S.D., 2002. Estimation of tropical forest structural
characteristics using large-footprint lidar. Remote Sens. Environ. 79, 305–319.
doi:10.1016/S0034-4257(01)00281-4

Duncanson, L.I., Dubayah, R.O., Cook, B.D., Rosette, J., Parker, G., 2015. The importance of
spatial detail: Assessing the utility of individual crown information and scaling
approaches for lidar-based biomass density estimation. Remote Sens. Environ. 168, 102–
112. doi:10.1016/j.rse.2015.06.021

Ene, L.T., Næsset, E., Gobakken, T., Gregoire, T.G., Ståhl, G., Nelson, R., 2012. Assessing
the accuracy of regional LiDAR-based biomass estimation using a simulation approach.
Remote Sens. Environ. 123, 579–592. doi:10.1016/j.rse.2012.04.017

Ewers, R.M., Didham, R.K., Fahrig, L., Ferraz, G., Hector, A., Holt, R.D., Turner, E.C., 2011.
A large-scale forest fragmentation experiment: the Stability of Altered Forest Ecosystems
Project. Philos. Trans. R. Soc. B 366, 3292–3302. doi:10.1098/rstb.2011.0049

Feldpausch, T.R., Banin, L., Phillips, O.L., Baker, T.R., Lewis, S.L., Quesada, C. a., Affum-
Baffoe, K., Arets, E.J.M.M., Berry, N.J., Bird, M., Brondizio, E.S., De Camargo, P.,
Chave, J., Djagbletey, G., Domingues, T.F., Drescher, M., Fearnside, P.M., França, M.B.,
Fyllas, N.M., Lopez-Gonzalez, G., Hladik, A., Higuchi, N., Hunter, M.O., Iida, Y., Salim,
K.A., Kassim, A.R., Keller, M., Kemp, J., King, D.A., Lovett, J.C., Marimon, B.S.,
Marimon-Junior, B.H., Lenza, E., Marshall, A.R., Metcalfe, D.J., Mitchard, E.T.A.,
Moran, E.F., Nelson, B.W., Nilus, R., Nogueira, E.M., Palace, M., Patiño, S., Peh, K.S.H.,
Raventos, M.T., Reitsma, J.M., Saiz, G., Schrodt, F., Sonké, B., Taedoumg, H.E., Tan, S.,
White, L., Wöll, H., Lloyd, J., 2011. Height-diameter allometry of tropical forest trees.
Biogeosciences 8, 1081–1106. doi:10.5194/bg-8-1081-2011

Fox, J.E.D., 1973. A handbook to Kabili-Sepilok Forest Reserve, Sabah Forest Record No. 9.
Borneo Literature Bureau, Kuching, Sarawak, Malaysia.

Gaveau, D.L.A., Sheil, D., Husnayaen, Salim, M.A., Arjasakusuma, S., Ancrenaz, M.,
Pacheco, P., Meijaard, E., 2016. Rapid conversions and avoided deforestation: examining
four decades of industrial plantation expansion in Borneo. Sci. Rep. 6, 32017.
doi:10.1038/srep32017

Gaveau, D.L.A., Sloan, S., Molidena, E., Yaen, H., Sheil, D., Abram, N.K., Ancrenaz, M.,
Nasi, R., Quinones, M., Wielaard, N., Meijaard, E., 2014. Four decades of forest
persistence, clearance and logging on Borneo. PLoS One 9, e101654.
doi:10.1371/journal.pone.0101654

Ghazoul, J., 2016. Dipterocarp Biology, Ecology, and Conservation. Oxford University Press.

doi:10.1093/acprof:oso/9780199639656.001.0001

Gibbs, H.K., Brown, S., Niles, J.O., Foley, J.A., 2007. Monitoring and estimating tropical forest carbon stocks: making REDD a reality. Environ. Res. Lett. 2, 45023. doi:10.1088/1748-9326/2/4/045023

Gobakken, T., Næsset, E., 2008. Assessing effects of laser point density, ground sampling intensity, and field sample plot size on biophysical stand properties derived from airborne laser scanner data. Can. J. For. Res. 38, 1095–1109. doi:10.1139/X07-219

Gonzalez, P., Asner, G.P., Battles, J.J., Lefsky, M.A., Waring, K.M., Palace, M., 2010. Forest carbon densities and uncertainties from Lidar, QuickBird, and field measurements in California. Remote Sens. Environ. 114, 1561–1575. doi:10.1016/j.rse.2010.02.011

Gregoire, T.G., Næsset, E., McRoberts, R.E., Ståhl, G., Andersen, H.-E., Gobakken, T., Ene, L., Nelson, R., 2016. Statistical rigor in LiDAR-assisted estimation of aboveground forest biomass. Remote Sens. Environ. 173, 98–108. doi:10.1016/j.rse.2015.11.012

Hansen, E., Gobakken, T., Bollandsås, O., Zahabu, E., Næsset, E., 2015. Modeling aboveground biomass in dense tropical submontane rainforest using airborne laser scanner data. Remote Sens. 7, 788–807. doi:10.3390/rs70100788

Hansen, M.C., Potapov, P. V, Moore, R., Hancher, M., Turubanova, S.A., Tyukavina, A., Thau, D., Stehman, S. V, Goetz, S.J., Loveland, T.R., Kommareddy, A., Egorov, A., Chini, L., Justice, C.O., Townshend, J.R.G., 2013. High-resolution global maps of 21st-century forest cover change. Science. 342, 850–3. doi:10.1126/science.1244693

HCS Approach, 2016. Public Consultation: HCS Approach Methodology; Toolkit version 2.0
| High Carbon Stock Approach.

HCS Approach Steering Group, 2015. The HSC Approach Toolkit. Kuala Lumpur.

Hector, A., Philipson, C., Saner, P., Chamagne, J., Dzulkifli, D., O'Brien, M., Snaddon, J.L.,
Ulok, P., Weilenmann, M., Reynolds, G., Godfray, H.C.J., 2011. The Sabah Biodiversity
Experiment: A long-term test of the role of tree diversity in restoring tropical forest
structure and functioning. Philos. Trans. R. Soc. B 366, 3303–3315.
doi:10.1098/rstb.2011.0094

Houghton, R.A., Byers, B., Nassikas, A.A., 2015. A role for tropical forests in stabilizing
atmospheric CO2. Nat. Clim. Chang. 5, 1022–1023. doi:10.1038/nclimate2869

Ioki, K., Tsuyuki, S., Hirata, Y., Phua, M.-H., Wong, W.V.C., Ling, Z.-Y., Saito, H., Takao,
G., 2014. Estimating above-ground biomass of tropical rainforest of different degradation
levels in Northern Borneo using airborne LiDAR. For. Ecol. Manage. 328, 335–341.
doi:10.1016/j.foreco.2014.06.003

Jubanski, J., Ballhorn, U., Kronseder, K., Franke, J., Siegert, F., J Franke, Siegert, F., 2013.
Detection of large above-ground biomass variability in lowland forest ecosystems by
airborne LiDAR. Biogeosciences 10, 3917–3930. doi:10.5194/bg-10-3917-2013

Kettle, C.J., Maycock, C.R., Burslem, D., 2012. New directions in dipterocarp biology and
conservation: a synthesis. Biotropica 44, 658–660. doi:10.1111/j.1744-
7429.2012.00912.x

Kumagai, T., Porporato, A., 2012. Drought-induced mortality of a Bornean tropical rain forest amplified by climate change. J. Geophys. Res. 117, 1–13. doi:10.1029/2011JG001835

Lamb, D., Erskine, P.D., Parrotta, J.A., 2005. Restoration of degraded tropical forest landscapes. Science. 310, 1628–1632.

Lefsky, M.A., Cohen, W.B., Parker, G.G., Harding, D.J., 2002. Lidar remote sensing for ecosystem studies. Bioscience 52, 19–30.

Lefsky, M., Cohen, W., Acker, S., Parker, G., 1999. Lidar remote sensing of the canopy structure and biophysical properties of Douglas-fir western hemlock forests. Remote Sens. Environ. 70, 339–361.

Longo, M., Keller, M.M., Dos-Santos, M.N., Leitold, V., Pinagé, E.R., Baccini, A., Saatchi, S., Nogueira, E.M., Batistella, M., Morton, D.C., 2016. Aboveground biomass variability across intact and degraded forests in the Brazilian Amazon. Global Biogeochem. Cycles 30, 1639–1660. doi:10.1002/2016GB005465

Malhi, Y., Wright, J., 2004. Spatial patterns and recent trends in the climate of tropical rainforest regions. Philos. Trans. R. Soc. B 359, 311–329. doi:10.1098/rstb.2003.1433

Manuri, S., Brack, C., Nugroho, N.P., Hergoualc'h, K., Novita, N., Dotzauer, H., Verchot, L., Putra, C.A.S., Widyasari, E., 2014. Tree biomass equations for tropical peat swamp forest ecosystems in Indonesia. For. Ecol. Manage. 334, 241–253. doi:10.1016/j.foreco.2014.08.031

Martin, A.R., Thomas, S.C., 2011. A reassessment of carbon content in tropical trees. PLoS
One 6, e23533. doi:10.1371/journal.pone.0023533

McRoberts, R.E., Næsset, E., Gobakken, T., 2016. The effects of temporal differences between
map and ground data on map-assisted estimates of forest area and biomass. Ann. For. Sci.
73, 839–847. doi:10.1007/s13595-015-0485-6

McRoberts, R.E., Næsset, E., Gobakken, T., 2013. Inference for lidar-assisted estimation of
forest growing stock volume. Remote Sens. Environ. 128, 268–275.
doi:10.1016/j.rse.2012.10.007

Mitchard, E.T.A., Feldpausch, T.R., Brienen, R.J.W., Lopez-Gonzalez, G., Monteagudo, A.,
Baker, T.R., Lewis, S.L., Lloyd, J., Quesada, C.A., Gloor, M., ter Steege, H., Meir, P.,
Alvarez, E., Araujo-Murakami, A., Aragão, L.E.O.C., Arroyo, L., Aymard, G., Banki, O.,
Bonal, D., Brown, S., Brown, F.I., Cerón, C.E., Chama Moscoso, V., Chave, J., Comiskey,
J.A., Cornejo, F., Corrales Medina, M., Da Costa, L., Costa, F.R.C., Di Fiore, A.,
Domingues, T.F., Erwin, T.L., Frederickson, T., Higuchi, N., Honorio Coronado, E.N.,
Killeen, T.J., Laurance, W.F., Levis, C., Magnusson, W.E., Marimon, B.S., Marimon
Junior, B.H., Mendoza Polo, I., Mishra, P., Nascimento, M.T., Neill, D., Núñez Vargas,
M.P., Palacios, W.A., Parada, A., Pardo Molina, G., Peña-Claros, M., Pitman, N., Peres,
C.A., Poorter, L., Prieto, A., Ramirez-Angulo, H., Restrepo Correa, Z., Roopsind, A.,
Roucoux, K.H., Rudas, A., Salomão, R.P., Schietti, J., Silveira, M., de Souza, P.F.,
Steininger, M.K., Stropp, J., Terborgh, J., Thomas, R., Toledo, M., Torres-Lezama, A.,
van Andel, T.R., van der Heijden, G.M.F., Vieira, I.C.G., Vieira, S., Vilanova-Torre, E.,

Vos, V.A., Wang, O., Zartman, C.E., Malhi, Y., Phillips, O.L., 2014. Markedly divergent estimates of Amazon forest carbon density from ground plots and satellites. Glob. Ecol. Biogeogr. 23, 935–946. doi:10.1111/geb.12168

Morales, J.L., Nocedal, J., 2011. Remark on Algorithm 778: L-BFGS-B, FORTRAN routines for large scale bound constrained optimization. ACM Trans. Math. Softw. 38, 1.

Nelson, R., Krabill, W., Tonelli, J., 1988. Estimating forest biomass and volume using airborne laser data. Remote Sens. Environ. 24, 247–267.

Nilus, R., Maycock, C., Majalap-Lee, N., Burslem, D., 2011. Nutrient limitation of tree seedling growth in three soil types found in Sabah. J. Trop. For. Sci. 23, 133–142.

Osman, R., Phua, M.-H., Ling, Z.Y., Kamlun, K.U., 2012. Monitoring of deforestation rate and trend in Sabah between 1990 and 2008 using multitemporal landsat data. J. For. Environ. Sci. 28, 144–151. doi:10.7747/JFS.2012.28.3.144

Pan, Y., Birdsey, R.A., Fang, J., Houghton, R.A., Kauppi, P.E., Kurz, W.A., Phillips, O.L., Shvidenko, A., Lewis, S.L., Canadell, J.G., Ciais, P., Jackson, R.B., Pacala, S.W., McGuire, A.D., Piao, S., Rautiainen, A., Sitch, S., Hayes, D., Canadell, J.G., Khatiwala, S., Primeau, F., Hall, T., Quéré, C. Le, Dixon, R.K., Kauppi, P.E., Kurz, W.A., Stinson, G., Rampley, G.J., Dymond, C.C., Neilson, E.T., Stinson, G., Birdsey, R.A., Pregitzer, K., Lucier, A., Kauppi, P.E., Pan, Y., Pan, Y., Birdsey, R.A., Hom, J., McCullough, K., Mantgem, P.J. van, Breshears, D.D., Ciais, P., Fang, J., Chen, A., Peng, C., Zhao, S., Ci, L., Lewis, S.L., Phillips, O.L., Gloor, M., Lewis, S.L., Lloyd, J., Sitch, S., Mitchard,

E.T.A., Laurance, W.F., Houghton, R.A., Friedlingstein, P., Tarnocai, C., Hooijer, A., Page, S.E., Rieley, J.O., Banks, C.J., McGuire, A.D., Goodale, C.L., Sarmiento, J.L., Schulze, E.D., Pacala, S.W., Phillips, O.L., Metsaranta, J.M., Kurz, W.A., Neilson, E.T., Stinson, G., Zhao, M., Running, S.W., Houghton, R.A., 2011. A large and persistent carbon sink in the world's forests. Science. 333, 988–93. doi:10.1126/science.1201609

Pfeifer, M., Kor, L., Nilus, R., Turner, E., Cusack, J., Lysenko, I., Khoo, M., Chey, V.K., Chung, A.C., Ewers, R.M., 2016. Mapping the structure of Borneo's tropical forests across a degradation gradient. Remote Sens. Environ. 176, 84–97. doi:10.1016/j.rse.2016.01.014

Phillips, O.L., Malhi, Y., Higuchi, N., Laurance, W.F., Núñez, P. V, Vásquez, R.M., Laurance, S.G., Ferreira, L. V, Stern, M., Brown, S., Grace, J., 1998. Changes in the carbon balance of tropical forests: Evidence from long-term plots. Science. 282, 439–442.

Popescu, S.C., Zhao, K., Neuenschwander, A., Lin, C., 2011. Satellite lidar vs. small footprint airborne lidar: Comparing the accuracy of aboveground biomass estimates and forest structure metrics at footprint level. Remote Sens. Environ. 115, 1–12. doi:10.1016/j.rse.2011.01.026

Putz, F.E., Zuidema, P.A., Synnott, T., Peña-Claros, M., Pinard, M.A., Sheil, D., Vanclay, J.K., Sist, P., Gourlet-Fleury, S., Griscom, B., Palmer, J., Zagt, R., 2012. Sustaining conservation values in selectively logged tropical forests: the attained and the attainable. Conserv. Lett. 5, 296–303. doi:10.1111/j.1755-263X.2012.00242.x

Quesada, C.A., Phillips, O.L., Schwarz, M., Czimczik, C.I., Baker, T.R., Patiño, S., Fyllas,
N.M., Hodnett, M.G., Herrera, R., Almeida, S., Alvarez Dávila, E., Arneth, A., Arroyo,
L., Chao, K.J., Dezzeo, N., Erwin, T., Di Fiore, A., Higuchi, N., Honorio Coronado, E.,
Jimenez, E.M., Killeen, T., Lezama, A.T., Lloyd, G., Löpez-González, G., Luizão, F.J.,
Malhi, Y., Monteagudo, A., Neill, D.A., Núñez Vargas, P., Paiva, R., Peacock, J., Peñuela,
M.C., Peña Cruz, A., Pitman, N., Priante Filho, N., Prieto, A., Ramírez, H., Rudas, A.,
Salomão, R., Santos, A.J.B., Schmerler, J., Silva, N., Silveira, M., Vásquez, R., Vieira, I.,
Terborgh, J., Lloyd, J., 2012. Basin-wide variations in Amazon forest structure and
function are mediated by both soils and climate. Biogeosciences 9, 2203–2246.
doi:10.5194/bg-9-2203-2012

R Core Development Team, 2016. R: A language and environment for statistical computing.
R Foundation for Statistical Computing, Vienna, Austria.

Réjou-Méchain, M., Muller-Landau, H.C., Detto, M., Thomas, S.C., Le Toan, T., Saatchi, S.S.,
Barreto-Silva, J.S., Bourg, N.A., Bunyavejchewin, S., Butt, N., Brockelman, W.Y., Cao,
M., Cárdenas, D., Chiang, J.-M.M., Chuyong, G.B., Clay, K., Condit, R., Dattaraja, H.S.,
Davies, S.J., Duque, A., Esufali, S., Ewango, C., Fernando, R.H.S.H.S., Fletcher, C.D.,
N. Gunatilleke, I.A.U., Hao, Z., Harms, K.E., Hart, T.B., Hérault, B., Howe, R.W.,
Hubbell, S.P., Johnson, D.J., Kenfack, D., Larson, A.J., Lin, L., Lin, Y., Lutz, J.A.,
Makana, J.-R.R., Malhi, Y., Marthews, T.R., McEwan, R.W., Mcmahon, S.M., Mcshea,
W.J., Muscarella, R., Nathalang, A., Noor, N.S.M.S.M., Nytch, C.J., Oliveira, A.A.,
Phillips, R.P., Pongpattananurak, N., Punchi-Manage, R., Salim, R., Schurman, J.,
Sukumar, R., Suresh, H.S., Suwanvecho, U., Thomas, D.W., Thompson, J., Uríarte, M.,

Valencia, R., Vicentini, A., Wolf, A.T., Yap, S., Yuan, Z., Zartman, C.E., Zimmerman, J.K., Chave, J., 2014. Local spatial structure of forest biomass and its consequences for remote sensing of carbon stocks. Biogeosciences 11, 6827–6840. doi:10.5194/bg-11-6827-2014

Réjou-Méchain, M., Tanguy, A., Piponiot, C., Chave, J., Hérault, B., 2017. BIOMASS: An R Package for estimating aboveground biomass and its uncertainty in tropical forests. Methods Ecol. Evol. doi:10.1111/2041-210X.12753

Réjou-Méchain, M., Tymen, B., Blanc, L., Fauset, S., Feldpausch, T.R., Monteagudo, A., Phillips, O.L., Richard, H., Chave, J., 2015. Using repeated small-footprint LiDAR acquisitions to infer spatial and temporal variations of a high-biomass Neotropical forest. Remote Sens. Environ. 169, 93–101. doi:10.1016/j.rse.2015.08.001

RSPO, 2007. Principles and Criteria for Sustainable Palm Oil Production.

Rutishauser, E., Noor'an, F., Laumonier, Y., Halperin, J., Rufi'ie, Hergoualc'h, K., Verchot, L., 2013. Generic allometric models including height best estimate forest biomass and carbon stocks in Indonesia. For. Ecol. Manage. 307, 219–225. doi:10.1016/j.foreco.2013.07.013

Saatchi, S.S., Harris, N.L., Brown, S., Lefsky, M., Mitchard, E.T.A., Salas, W., Zutta, B.R., Buermann, W., Lewis, S.L., Hagen, S., Petrova, S., White, L., Silman, M., Morel, A., 2011. Benchmark map of forest carbon stocks in tropical regions across three continents. Proc. Natl. Acad. Sci. 108, 9899–9904. doi:10.1073/pnas.1019576108

Saner, P., Loh, Y.Y., Ong, R.C., Hector, A., 2012. Carbon stocks and fluxes in tropical lowland
dipterocarp rainforests in Sabah, Malaysian Borneo. PLoS One 7, e29642.
doi:10.1371/journal.pone.0029642

Särndal, C.-E., Swensson, B., Wretman, J.H., 1992. Model assisted survey sampling. Springer-
Verlag.

Singh, M., Evans, D., Coomes, D.A., Friess, D.A., Suy Tan, B., Samean Nin, C., 2016.
Incorporating canopy cover for airborne-derived assessments of forest biomass in the
tropical forests of Cambodia. PLoS One 11, e0154307. doi:10.1371/journal.pone.0154307

Slik, J.W.F., Aiba, S.I., Brearley, F.Q., Cannon, C.H., Forshed, O., Kitayama, K., Nagamasu,
H., Nilus, R., Payne, J., Paoli, G., Poulsen, A.D., Raes, N., Sheil, D., Sidiyasa, K., Suzuki,
E., van Valkenburg, J.L.C.H., 2010. Environmental correlates of tree biomass, basal area,
wood specific gravity and stem density gradients in Borneo's tropical forests. Glob. Ecol.
Biogeogr. 19, 50–60. doi:10.1111/j.1466-8238.2009.00489.x

Slik, J.W.F., Bernard, C.S., Breman, F.C., Van Beek, M., Salim, A., Sheil, D., 2008. Wood
density as a conservation tool: quantification of disturbance and identification of
conservation-priority areas in tropical forests. Conserv. Biol. 22, 1299–1308.
doi:10.1111/j.1523-1739.2008.00986.x

Spriggs, R., 2015. Robust methods for estimating forest stand characteristics across landscapes
using airborne LiDAR. University of Cambridge.

Struebig, M.J., Turner, A., Giles, E., Lasmana, F., Tollington, S., Bernard, H., Bell, D., 2013.

Quantifying the biodiversity value of repeatedly logged rainforests: gradient and comparative approaches from Borneo. Adv. Ecol. Res. 48, 183–224. doi:10.1016/B978-0-12-417199-2.00003-3

Struebig, M.J., Wilting, A., Gaveau, D.L.A., Meijaard, E., Smith, R.J., Abdullah, T., Abram, N., Alfred, R., Ancrenaz, M., Augeri, D.M., Belant, J.L., Bernard, H., Bezuijen, M., Boonman, A., Boonratana, R., Boorsma, T., Breitenmoser-Würsten, C., Brodie, J., Cheyne, S.M., Devens, C., Duckworth, J.W., Duplaix, N., Eaton, J., Francis, C., Fredriksson, G., Giordano, A.J., Gonner, C., Hall, J., Harrison, M.E., Hearn, A.J., Heckmann, I., Heydon, M., Hofer, H., Hon, J., Husson, S., Anwarali Khan, F.A., Kingston, T., Kreb, D., Lammertink, M., Lane, D., Lasmana, F., Liat, L.B., Lim, N.T.-L., Lindenborn, J., Loken, B., Macdonald, D.W., Marshall, A.J., Maryanto, I., Mathai, J., McShea, W.J., Mohamed, A., Nakabayashi, M., Nakashima, Y., Niedballa, J., Noerfahmy, S., Persey, S., Peter, A., Pieterse, S., Pilgrim, J.D., Pollard, E., Purnama, S., Rafiastanto, A., Reinfelder, V., Reusch, C., Robson, C., Ross, J., Rustam, R., Sadikin, L., Samejima, H., Santosa, E., Sapari, I., Sasaki, H., Scharf, A.K., Semiadi, G., Shepherd, C.R., Sykes, R., van Berkel, T., Wells, K., Wielstra, B., Wong, A., Fischer, M., Metcalfe, K., Kramer-Schadt, S., 2015. Targeted conservation to safeguard a biodiversity hotspot from climate and land-cover change. Curr. Biol. 25, 372–378. doi:10.1016/j.cub.2014.11.067

Sullivan, M.J.P., Talbot, J., Lewis, S.L., Phillips, O.L., Qie, L., Begne, S.K., Chave, J., Cuni-Sanchez, A., Hubau, W., Lopez-Gonzalez, G., Miles, L., Monteagudo-Mendoza, A., Sonké, B., Sunderland, T., ter Steege, H., White, L.J.T., Affum-Baffoe, K., Aiba, S., de

Almeida, E.C., de Oliveira, E.A., Alvarez-Loayza, P., Dávila, E.Á., Andrade, A., Aragão, L.E.O.C., Ashton, P., Aymard C., G.A., Baker, T.R., Balinga, M., Banin, L.F., Baraloto, C., Bastin, J.-F., Berry, N., Bogaert, J., Bonal, D., Bongers, F., Brienen, R., Camargo, J.L.C., Cerón, C., Moscoso, V.C., Chezeaux, E., Clark, C.J., Pacheco, Á.C., Comiskey, J.A., Valverde, F.C., Coronado, E.N.H., Dargie, G., Davies, S.J., De Canniere, C., Djuikouo K., M.N., Doucet, J.-L., Erwin, T.L., Espejo, J.S., Ewango, C.E.N., Fauset, S., Feldpausch, T.R., Herrera, R., Gilpin, M., Gloor, E., Hall, J.S., Harris, D.J., Hart, T.B., Kartawinata, K., Kho, L.K., Kitayama, K., Laurance, S.G.W., Laurance, W.F., Leal, M.E., Lovejoy, T., Lovett, J.C., Lukasu, F.M., Makana, J.-R., Malhi, Y., Maracahipes, L., Marimon, B.S., Junior, B.H.M., Marshall, A.R., Morandi, P.S., Mukendi, J.T., Mukinzi, J., Nilus, R., Vargas, P.N., Camacho, N.C.P., Pardo, G., Peña-Claros, M., Pétronelli, P., Pickavance, G.C., Poulsen, A.D., Poulsen, J.R., Primack, R.B., Priyadi, H., Quesada, C.A., Reitsma, J., Réjou-Méchain, M., Restrepo, Z., Rutishauser, E., Salim, K.A., Salomão, R.P., Samsoedin, I., Sheil, D., Sierra, R., Silveira, M., Slik, J.W.F., Steel, L., Taedoumg, H., Tan, S., Terborgh, J.W., Thomas, S.C., Toledo, M., Umunay, P.M., Gamarra, L.V., Vieira, I.C.G., Vos, V.A., Wang, O., Willcock, S., Zemagho, L., 2017. Diversity and carbon storage across the tropical forest biome. Sci. Rep. 7, 39102. doi:10.1038/srep39102

Vaglio Laurin, G., Chen, Q., Lindsell, J.A., Coomes, D.A., Frate, F. Del, Guerriero, L., Pirotti, F., Valentini, R., 2014. Above ground biomass estimation in an African tropical forest with lidar and hyperspectral data. ISPRS J. Photogramm. Remote Sens. 89, 49–58. doi:10.1016/j.isprsjprs.2014.01.001

Vincent, G., Sabatier, D., Rutishauser, E., 2014. Revisiting a universal airborne light detection and ranging approach for tropical forest carbon mapping: scaling-up from tree to stand to landscape. Oecologia 175, 439–43. doi:10.1007/s00442-014-2913-y

Vira, B., Christoph, W., Mansourian, S., 2015. Forests, Trees and Landscapes for Food Security and Nutrition - A Global Assessment Report. Vienna.

Walsh, R.P.D., Newbery, D.M., 1999. The ecoclimatology of Danum, Sabah, in the context of the world's rainforest regions, with particular reference to dry periods and their impact. Philos. Trans. R. Soc. B 354, 1869–83. doi:10.1098/rstb.1999.0528

Wulder, M.A., White, J.C., Nelson, R.F., Næsset, E., Ørka, H.O., Coops, N.C., Hilker, T., Bater, C.W., Gobakken, T., 2012. Lidar sampling for large-area forest characterization: A review. Remote Sens. Environ. 121, 196–209. doi:10.1016/j.rse.2012.02.001

Yanai, R.D., Battles, J.J., Richardson, A.D., Blodgett, C.A., Wood, D.M., Rastetter, E.B., 2010. Estimating uncertainty in ecosystem budget calculations. Ecosystems 13, 239–248. doi:10.1007/s10021-010-9315-8

Zanne, A.E., Lopez-Gonzalez, G., Coomes, D.A., Ilic, J., Jansen, S., Lewis, S.L., Miller, R.B., Swenson, N.G., Wiemann, M.C., Chave, J., 2009. Global wood density database. Dryad Digit. Repos. http//dx.doi.org/10.5061/dryad.234.

Zolkos, S.G., Goetz, S.J., Dubayah, R., 2013. A meta-analysis of terrestrial aboveground biomass estimation using lidar remote sensing. Remote Sens. Environ. 128, 289–298. doi:10.1016/j.rse.2012.10.017

# Supporting information

**Appendix S1 |** Quantifying aboveground carbon density and its uncertainty

*Correcting stem diameters for position of measurement*

Stem diameters ($D$, in cm) are typically recorded at a height of 1.3 m aboveground. However, in some cases it may be necessary to measure $D$ at a different point along the stem (e.g., in the presence of stem deformities or buttress roots). To account for differences in the position of measurement ($POM$, in m), we used the following taper model developed for Neotropical forests by Cushman *et al*. (2014) to reconstruct stem diameters at a height of 1.3 m aboveground ($D_{1.3m}$):

$$D_{1.3m} = \frac{D_{POM}}{\exp\big(-0.029 \times (POM - 1.3)\big)} \tag{S1}$$

where $D_{POM}$ is the stem diameter measurement taken at $POM$, which in in turn is expressed as a height in meters aboveground. When not reported, $POM$ was assumed to be at 1.3 m aboveground.

*Tree height estimation*

Tree heights ($H$, in m) were measured for a subset of trees at Sepilok ($n = 718$), at Kuamut ($n = 5587$), in the SAFE experimental plots ($n = 7653$) and in the riparian buffer zones within the SAFE landscape ($n = 1380$), in the Global Ecosystem Monitoring (GEM) plots ($n = 2769$), and in both the CTSF plot and the CAO plots established at Danum Valley ($n = 836$ and $n = 2769$, respectively). In each case, $H$ was measured using a laser range finder. Using these data, we developed site-specific $H$–$D$ equations in order to estimate the height of trees that were not measured. Following the protocol outlined in the *BIOMASS* package in R (Rejou-Mechain et al., 2016), we compared a number of alternative *H–D* models with the intent of minimizing the residual standard error (σ) of the model. We found that a mixed-effects model of the form $\ln(H) = \rho_0 + \rho_1 \times \ln(D) + \rho_2 \times \ln(D)^2$, where $\rho_{0–2}$ were allowed to vary by site (i.e., site was treated as a random effect influencing both the intercept and slope of the model), fit the data best (σ = 4.4; $R^2$ = 0.84). Fig. S1 illustrates the fit of this equation to the data.

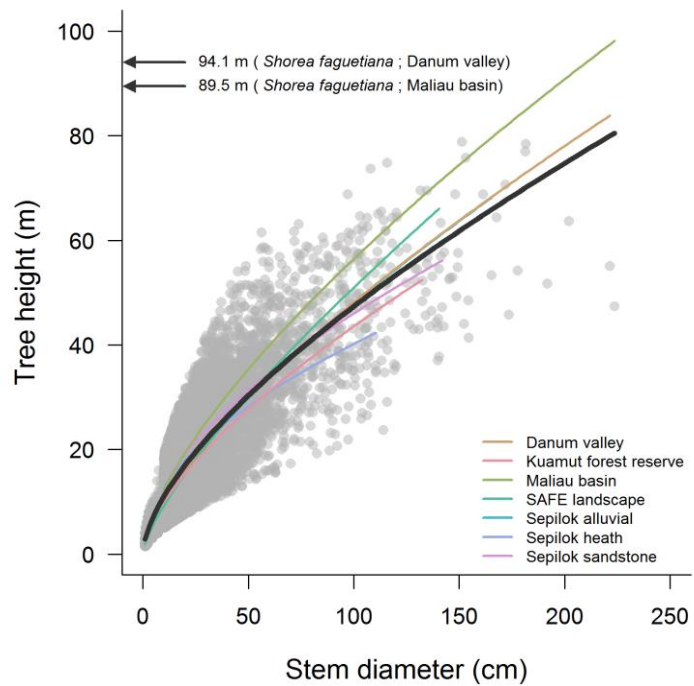

**Fig. S1** | Relationship between tree height and stem diameter across study sites in Sabah. The black curve corresponds to the best fit *H–D* equation across all sites, while coloured lines illustrate how *H–D* relationships vary among sites. The height of the tallest known tree in Sabah (a 94.1 m tall *Shorea faguetiana* growing at Danum Valley, currently the tallest known tree in the tropics) and of the tallest tree at Maliau Basin (an 89.5 m tall *Shorea faguetiana*) are shown for context.

*Wood density estimation*

Wood density (*WD*, in g cm$^{-3}$) values were obtained from the Global Wood Density Database (Chave et al., 2009; Zanne et al., 2009). Prior to assigning *WD*, we first checked species names against those in the Taxonomic Name Resolution Service (Boyle et al., 2013). At Sepilok, in the CTSF plot at Danum Valley, in the GEM plots and in the CAO plots established at Danum Valley and Kuamut, trees were matched to species or closest taxonomic unit. If no taxonomic information was available, the mean *WD* of the plot was used instead (Talbot *et al.*, 2014; Rejou-Mechain *et al.*, 2016). For plots established as part of the SAFE forest fragmentation experiment and those in riparian buffer zones, trees were not identified taxonomically. In this case, *WD* values were assigned on the basis of disturbance history using data from the GEM plots as reference (see Table S1 for details).

**Table S1 |** Wood density values assigned to plots in the SAFE forest fragmentation experiment and in riparian buffer zones. For a description of the SAFE project, including the layout of the experimental blocks see Ewers *et al.* (2011). For a description of Global Ecosystem Monitoring (GEM) network plots see http://gem.tropicalforests.ox.ac.uk.

| Site | Plot type | SAFE block | Disturbance history | GEM plot code | Wood density (g cm$^{-3}$) |
|------|-----------|------------|---------------------|---------------|---------------|
| Maliau Basin | SAFE experiment | OG1, OG2  and OG3 | Old growth forest | Belian and Seraya | 0.57 |
| SAFE landscape | SAFE experiment | VJR | Low logging intensity | Belian and Seraya | 0.57 |
| SAFE landscape | SAFE experiment | LFE, LF1,  LF2 and LF3 | Twice-logged continuous forest | LFE | 0.61 |
| SAFE landscape | SAFE experiment | B | Twice-logged fragmented forest | B north and B south | 0.46 |
| SAFE landscape | SAFE experiment | E | Twice-logged fragmented forest | E | 0.53 |
| SAFE landscape | SAFE experiment | A, C, D and F | Twice-logged fragmented forest | E, B north and B south | 0.48 |
| SAFE landscape | Riparian buffers | LFE | Twice-logged continuous forest | LFE | 0.61 |
| SAFE landscape | Riparian buffers |  | Twice-logged fragmented forest | E, B north and B south | 0.48 |

*Accounting for missing stems in ACD and basal area estimation*

With the exception of plots at Danum Valley and 38 of the SAFE experimental plots where all stems >1 cm in *D* were recorded, in other datasets compiled for this study the size threshold for inclusion was *D* = 10 cm. While large trees account for most of the biomass in forests (e.g., Bastin *et al*., 2015), excluding small stems will nonetheless result in an underestimation of aboveground carbon density (*ACD*, in Mg C ha$^{-1}$), as well as basal area (*BA*, in m$^2$ ha$^{-1}$). To correct for this, we used the 45 1 ha plots at Danum to calculate *ACD* and *BA* using all available data (*ACD$_{1cm}$* and *BA$_{1cm}$*), and again after having excluded stems with *D* < 10 cm (*ACD$_{10cm}$* and *BA$_{10cm}$*). These data were then used to derive the following correction factors for *ACD* and *BA* which were applied to all other datasets (Fig. S2):

$$ACD_{1cm} = 6.713 + 1.004 \times ACD_{10cm} \quad\quad (S2)$$

$$BA_{1cm} = 4.168 + 1.009 \times BA_{10cm} \quad\quad (S3)$$

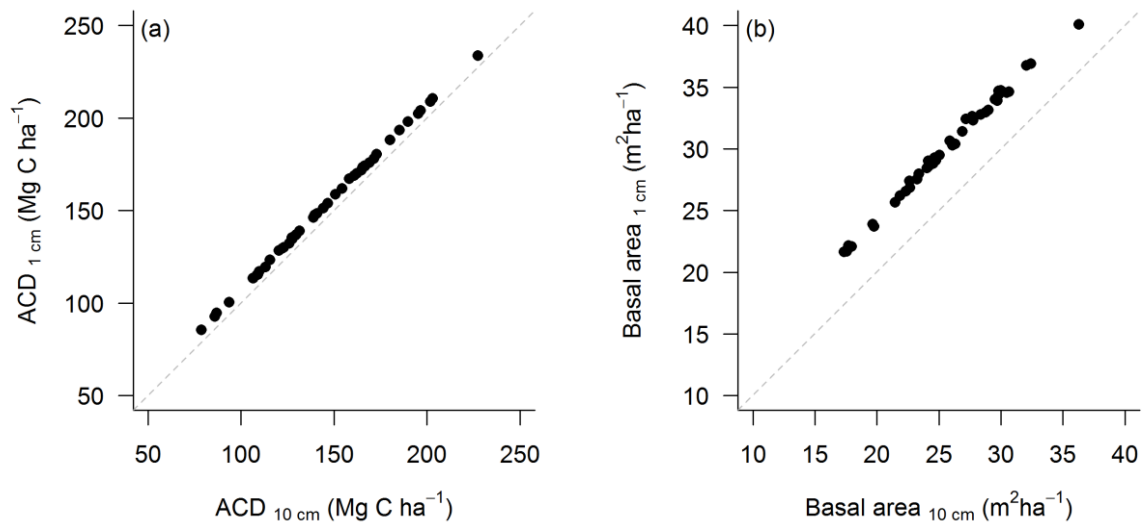

**Fig. S2** | Relationship between (**a**) aboveground carbon density (*ACD*) and (**b**) basal area (*BA*) calculated with all stems > 1 cm in diameter (*D*) and after excluding stems with $D < 10$ cm for the 45 1 ha plots at Danum Valley. Dashed lines correspond to a 1:1 relationship.

**Appendix S2 |** Basal area and wood density predictions from ALS

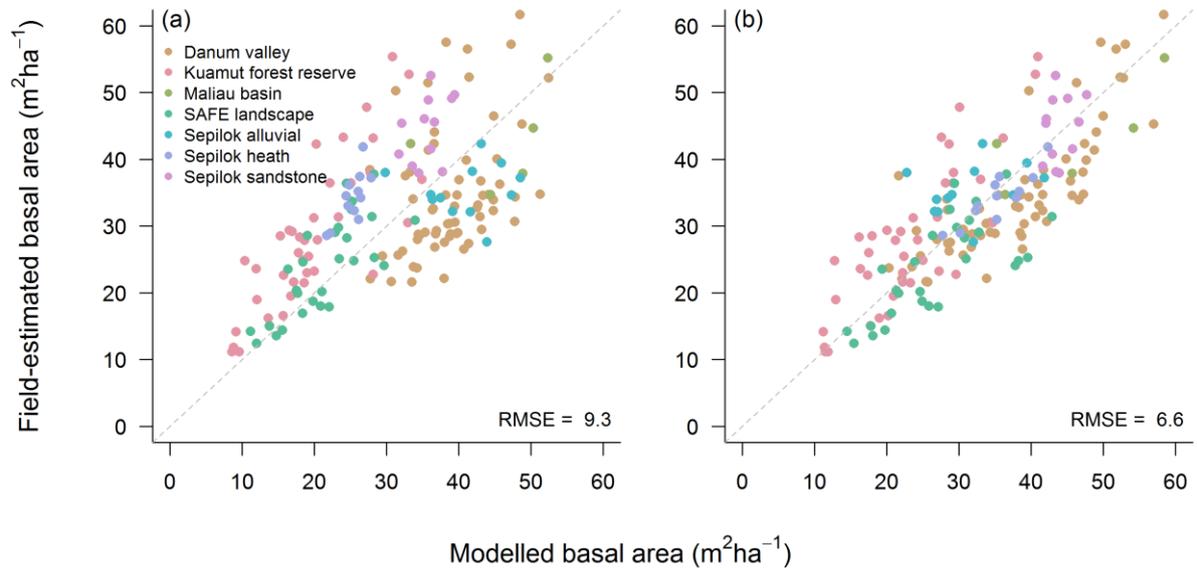

**Fig. S3 |** Relationship between field-estimated basal area (*BA*) and *BA* modelled as a function of (**a**) top-of-canopy height (*TCH*) [Eq. (9) in Table S2] and (**b**) a combination of *TCH* and canopy cover at 20 m aboveground [Eq. (10) in Table S2]. Dashed lines correspond to a 1:1 relationship. The RMSE of each comparison is printed in the bottom right-hand corner of the panels.

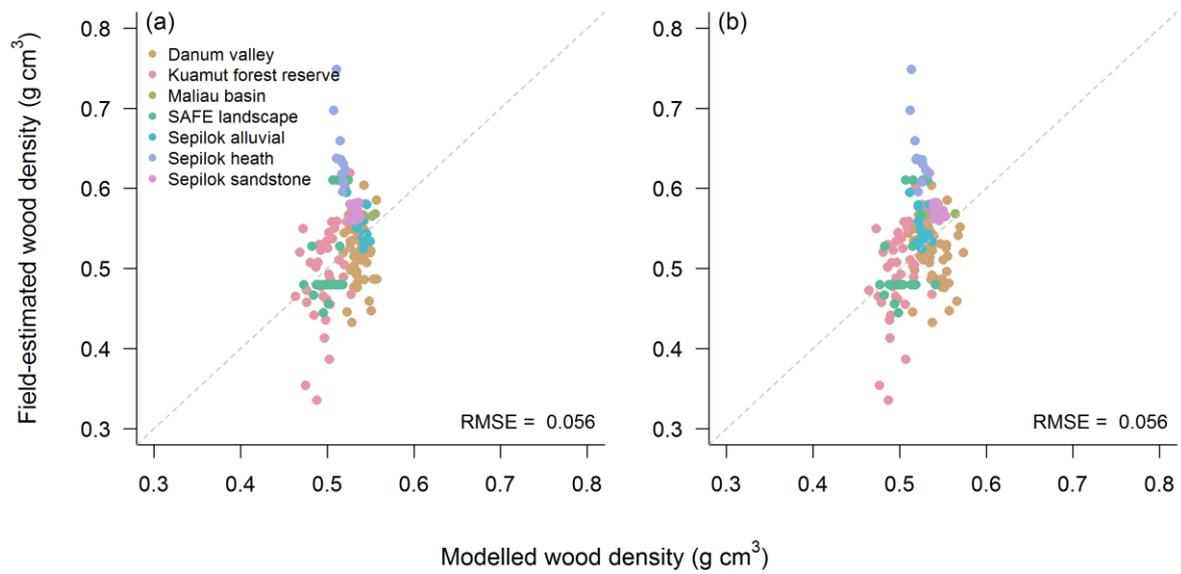

**Fig. S4 |** Relationship between field-estimated community-weighted mean wood density (*WD*) and *WD*
modelled as a function of (**a**) top-of-canopy height [Eq. (11) in Table S2] and (**b**) canopy cover at 20 m
aboveground. Dashed lines correspond to a 1:1 relationship. The RMSE of each comparison is printed
in the bottom right-hand corner of the panels.

**Appendix S3 | Confidence intervals for parameter estimates**

**Table S2 |** Parameter estimates for models presented in the main text. For Eq. (8–11), best-fit parameter estimates are mean values across 100 model iterations, with 95% confidence intervals reflecting variation across all 100 model runs. For each model, the equation number corresponds to that in the main text. $\sigma$ is the residual standard error of the model.

| Eq. | Model | $\rho_0$ | $\rho_1$ | $\rho_2$ | $\rho_3$ | $\sigma$ |
|---|---|---|---|---|---|---|
| 4 | $ln\left(\frac{Cover_{20}}{1 - Cover_{20}}\right) = \rho_0 + \rho_1 \times ln(TCH)$ | -12.431 | 4.061 | | | 0.101 |
| 8 | $ACD = \rho_0 \times TCH^{\rho_1} \times BA^{\rho_2} \times WD^{\rho_3}$ | 0.567 [0.389; 0.829] | 0.554 [0.452; 0.657] | 1.081 [0.956; 1.213] | 0.186 [-0.017; 0.351] | 0.185 |
| 9 | $BA = \rho_0 \times TCH$ | 1.112 [1.084; 1.142] | | | | 9.393 |
| 10 | $BA = \rho_0 \times TCH^{\rho_1} \times (1 + \rho_2 \times Cover_{resid})$ | 1.287 [1.217; 1.464] | 0.987 [0.945; 1.000] | 1.983 [1.904; 2.000] | | 6.581 |
| 11 | $WD = \rho_0 \times TCH^{\rho_1}$ | 0.385 [0.279; 0.516] | 0.097 [-0.013; 0.216] | | | 0.225 |

**Appendix S4 |** Comparison with spatial regression models

We tested for spatial autocorrelation (SAC) in the residuals of the regionally-calibrated *ACD* model [Eq. (8) in Table S2] by comparing it to a model with spatially correlated errors. Specifically, we used generalized least squares (GLS) regression to fit a model in which errors were assumed to be spatially correlated following an exponential decay function of geographic distance (Crawley, 2007). A semivariogram plot constructed using the residuals of the two models revealed little evidence of SAC for both the spatially naïve and the spatially explicit model (Fig. S5). Consequently, we chose to focus on the non-spatial model for the purpose of estimating *ACD*.

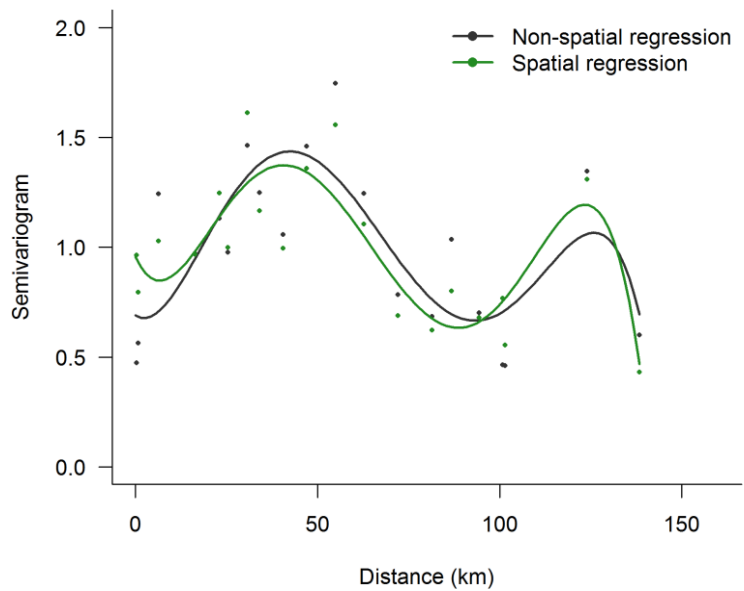

**Fig. S5 |** Semivariogram showing the degree of spatial autocorrelation for spatial and non-spatial models of aboveground carbon density [Eq. (8) in Table S2].

# References

Bastin J-F, Barbier N, Réjou-Méchain M et al. (2015) Seeing Central African forests through their largest trees. *Scientific Reports*, **5**, 13156.

Boyle B, Hopkins N, Lu Z et al. (2013) The taxonomic name resolution service: an online tool for automated standardization of plant names. *BMC bioinformatics*, **14**, 16.

Chave J, Coomes DA, Jansen S, Lewis SL, Swenson NG, Zanne AE (2009) Towards a worldwide wood economics spectrum. *Ecology Letters*, **12**, 351–366.

Crawley MJ (2007) *The R book*. Wiley, Chichester, England.

Cushman KC, Muller-Landau HC, Condit RS, Hubbell SP (2014) Improving estimates of biomass change in buttressed trees using tree taper models. *Methods in Ecology and Evolution*, **5**, 573–582.

Ewers RM, Didham RK, Fahrig L, Ferraz G, Hector A, Holt RD, Turner EC (2011) A large-scale forest fragmentation experiment: the Stability of Altered Forest Ecosystems Project. *Philosophical Transactions of the Royal Society B*, **366**, 3292–3302.

Rejou-Mechain M, Tanguy A, Piponiot C, Chave J, Herault B (2016) *BIOMASS: Estimating above-ground biomass and its uncertainty in tropical forests*. R package version 1.0. https://CRAN.R-project.org/package=BIOMASS.

Talbot J, Lewis SL, Lopez-Gonzalez G et al. (2014) Methods to estimate aboveground wood productivity from long-term forest inventory plots. *Forest Ecology and Management*, **320**, 30–38.

Zanne AE, Lopez-Gonzalez G, Coomes DA et al. (2009) Global wood density database. *Dryad Digital Repository, http://dx.doi.org/10.5061/dryad.234*.